\documentclass[prx,longbibliography,
twocolumn,showkeys,showpacs,superscriptaddress,nofootinbib,10pt,floatfix]{revtex4-1}

\usepackage{graphicx}
\usepackage{grffile}
\usepackage{amsthm}
\usepackage{epsfig}
\usepackage{amsmath,amssymb,amsfonts,mathtools,bbm,MnSymbol,mathrsfs,xfrac}
\usepackage{longtable}
\usepackage{tabularx}
\usepackage{colortbl}
\usepackage[table]{xcolor}
\usepackage[breaklinks=true,colorlinks=true,linkcolor=blue,urlcolor=blue,citecolor=blue]{hyperref}
\usepackage{color}
\usepackage[makeroom]{cancel}

\usepackage{soul}
\pdfoutput=1

\renewcommand{\[}{\begin{equation}}
\renewcommand{\]}{\end{equation}}

\newcommand{\ket}[1]{|#1\rangle}

\newcommand{\pro}[2]{|#1\rangle\langle#2|}
\newcommand{\mean}[1]{\langle#1\rangle}

\newcommand{\tr}{\mathrm{tr}}
\renewcommand{\th}{{\mathrm{th}}}

\newcommand{\R}{{\hat{\rho}}}

\newcommand{\C}{{\mathcal{C}}}
\renewcommand{\P}{\hat{P}}
\newcommand{\bi}{{\boldsymbol{i}}}

\newcommand{\HS}{\mathcal{H}}

\newcommand{\DE}{{\Delta\! E}}

\newcommand{\bx}{{\boldsymbol{x}}}
\newcommand{\bp}{{\boldsymbol{p}}}
\newcommand{\bn}{{\boldsymbol{n}}}
\newcommand{\bE}{{\boldsymbol{E}}}

\definecolor{mygray}{gray}{0.6}

\theoremstyle{definition}

\definecolor{dfcol}{cmyk}{1, 0.2108, 0.13, 0.3}
\newcommand{\df}[1]{\ifthenelse{\boolean{}}{\textcolor{dfcol}{[{\bf DF}: #1]}}{}}

\begin{document}

% title
\title{A brief introduction to observational entropy
%, a framework for equilibrium and non-equilibrium, classical and quantum, coarse-grained entropy
}

% authors

\author{Dominik \v{S}afr\'{a}nek}
\email{dsafranekibs@gmail.com}
\affiliation{SCIPP and Department of Physics, University of California, Santa Cruz, California 95064, USA}
\affiliation{Center for Theoretical Physics of Complex Systems, Institute for Basic Science (IBS), Daejeon 34126, Republic of Korea}

\author{Anthony Aguirre}
\affiliation{SCIPP and Department of Physics, University of California, Santa Cruz, California 95064, USA}

\author{Joseph Schindler}
\affiliation{SCIPP and Department of Physics, University of California, Santa Cruz, California 95064, USA}

\author{J. M. Deutsch}
\affiliation{Department of Physics, University of California, Santa Cruz, CA 95064, USA}

% date
\date{Sep 2021}

% abstract
\begin{abstract}
In the past several years, observational entropy has been developed as both a (time-dependent) quantum generalization of Boltzmann entropy, and as a rather general framework to encompass classical and quantum equilibrium and non-equilibrium coarse-grained entropy. In this paper we review the construction, interpretation, most important properties, and some applications of this framework. The treatment is self-contained and relatively pedagogical, aimed at a broad class of researchers.
\end{abstract}

% pacs and keys

\maketitle

\section{Introduction}

If you ask a working physicist ``what is energy?'' you are likely to get a reply close to ``a conserved quantity associated with the time-translation invariance of the laws of physics.'' But if you ask what the --- arguably equally fundamental --- concept of {\em entropy} means, you are likely to receive a bizarrely diverse set of answers.  It is a thermodynamic quantity related to heat transfer and temperature. It quantifies the genericity of the set of macroscopic properties that a system has. It is a measure of the information in a system.  It is a measure of the {\em uncertainty} in a system. It measures the quantum correlations between one part of a system and another. It is a quantity that increases in a closed system, per the second law of thermodynamics, underlying the arrow of time. It is one fourth of the area of an event horizon, in Planck units. And so on, each with a fairly distinct mathematical definition. These notions are certainly related, and some of the relations are fairly  clear; but some are quite obscure or ambiguous, both conceptually and mathematically.\footnote{For example how does a system in a pure state have both zero von Neumann entropy and nonzero thermodynamic entropy? Does black hole entropy correspond to entanglement, or coarse-graining? If information is preserved in a closed system, and entropy is information, how does entropy increase? etc.} 

Over the past several years the authors and others have developed the framework of {\em observational entropy}~\cite{safranek2019short,safranek2019long,safranek2019classical,strasberg2019entropy,faiez2020typical,strasberg2020heat,schindler2020entanglement,riera2020quantum}. Starting as a quantum version of Boltzmann entropy,\footnote{A generalization of Boltzmann entropy to quantum systems was first proposed by von Neumann citing personal discussion with Eugene Wigner~\cite{von2010proof}. He did this after expressing dissatisfaction with the von Neumann entropy as a proper measure of thermodynamic entropy, since it is ``computed from the perspective of an observer who can carry out all measurements that are possible in principle, i.e., regardless of whether they are macroscopic (for example, there every pure state has entropy 0, only mixtures have entropies greater than 0!).'' Since then the concept, also called ``coarse-grained'' entropy has appeared in literature both in quantum~\cite{von1955mathematical,wehrl1978general,gemmer2014entropy,almheiri2020entropy} and classical~\cite{wehrl1978general,latora1999kolmogorov,nauenberg2004evolution,kozlov2007fine,piftankin2008gibbs,vzupanovic2018relation} systems, but has been studied systematically only very recently.} observational entropy has evolved into a way to mathematically and conceptually unify many of these disparate concepts.  Given a system's state space, a probability density over this space, and one or more coarse-grainings of the space into distinct measurement outcomes,
an observer could obtain knowledge of the system by performing the measurement.
Observational entropy corresponds to the uncertainty in, i.e. lack of, this knowledge.

This framework is general enough to include both classical and quantum systems (and even more general ones), and also to correspond to many other entropies as special cases. If, for example, the coarse-graining is really a ``fine-graining'' into individual states, observational entropy can become Gibbs or von Neumann entropy.  If the measurements can access only part of a multipartite system, observational entropy can be used to define a generalization of entanglement entropy. If the coarse-graining is in energy, observational entropy corresponds to equilibrium thermodynamic entropy, and with further localized coarse graining in position also provides a definition of {\em non}-equilibrium thermodynamic entropy. By including a bath that is coarse-grained over, the framework can also be applied to open systems.

Coarse-graining is, of course, a very widely used concept, and while relatively novel when considered in full, the framework uses much of the same coarse-graining formalism that is sometimes precisely and sometimes loosely defined and used in the other fields of physics including the consistent histories quantum theory~\cite{gellmann1993classical,dowker1996on,griffiths2019consistent}, Kolmogorov-Sinai entropy~\cite{farmer1982information,latora1999kolmogorov,frigg2004sense,jost2006dynamical}, topological entropy~\cite{farmer1982information,jost2006dynamical}, entropy of an observable/entropy of partition~\cite{daniel1984entropy,jost2006dynamical,anza2017information,lent2019quantum,goldstein2019gibbs}, Black holes~\cite{engelhardt2019coarse,almheiri2020entropy}, and coarse-grained free energies (with applications in fluid dynamics~\cite{espanol1997coarse,gao2017analytical,batchelor1967introduction},
chemical engineering~\cite{smith1950introduction,guggenheim1956statistical,callen1998thermodynamics}, statistical mechanics of fields and renormalization group~\cite{fisher1998renormalization,kardar2007statisticalparticles,kardar2007statisticalfields,ma2018modern}, and field theory in the guise of renormalization~\cite{wilson1971renormalization}). 

The aim of this paper is to provide a concise but fairly complete treatment of the observational entropy framework and its physical motivations, along with some of the main results of the current state-of-the-art.
We will start with the quantum version in Sec.~\ref{sec-qm} with a single coarse-graining, then generalize to multiple coarse-grainings and classical physics in the next two sections. In Sec.~\ref{sec-lcg} we define local coarse-grainings and connect with entanglement, and in Sec.\ref{sec-phys} apply the framework to concrete example systems and explicitly construct coarse-grainings yielding equilibrium and non-equilibrium thermodynamic entropy.  In the concluding Sec.\ref{sec-conc}, we summarize and point out some open questions and directions forward.

\section{Construction}
\label{sec-qm}
Let us assume that the Hilbert space can be decomposed into a direct sum of orthogonal subspaces $\HS=\bigoplus_i\HS_i$, where each subspace corresponds to a \emph{macrostate} specifying a single macroscopic property of the system (such as energy or number of particles).\footnote{Some authors~\cite{gellmann1993classical} simply use the term ``property'' synonymously with ``macrostate,'' which is a useful conceptualization.} Defining $\P_i$ as the projector onto a subspace $\HS_i$, the set $\C=\{\P_i\}$ forms a set of Hermitian ($\P_i^\dag=\P_i$) orthogonal ($\P_i \P_j = \P_i \delta_{ij}$) projectors that form a partition of unity ($\sum_i\P_i=\hat{I}$), termed a \emph{coarse-graining}. Since a macroscopic property determined by a measuring apparatus is described by an observable, a natural way to specify a coarse-graining is via the spectral decomposition of an observable operator $\hat{A} = \sum_a a \, \P_{a}$ (each $a$ assumed to be distinct), with associated coarse-graining $\C_{\hat{A}} = \{ \P_a \}$.

The probability that a quantum state $\R$ will be found in a given macrostate can be calculated as $p_i=\tr[\P_i\R]$. Equivalently, we can say that this is the probability that a system described by a quantum state $\R$ will be found to have value $i$ of a macroscopic property, when performing a coarse-grained measurement on it, in the basis given by the coarse-graining.

Now, following Boltzmann's original conception, if we assume that an observer cannot distinguish between different microstates $k$ within the same macrostate $i$, we may associate the same probability $p_{i,k}=p_i/V_i$ to every microstate (given by a pure quantum state) in the macrostate, where $V_i=\dim(\HS_i)=\tr[\P_i]$ is the number of orthogonal pure states that fit into the macrostate; we call $V_i$ \emph{volume} of the macrostate. With this assignment, the statistical entropy that the observer associates with the system can be taken be the Shannon entropy of probabilities the $p_{i,k}$:
\[\label{eq:shannon}
-\sum_{i,k} p_{i,k}\ln p_{i,k}.
\]
This, after inserting $p_{i,k}=p_i/V_i$, reduces to
\[
\label{eq:def1}
S_{\C}\equiv-\sum_{i}p_i\ln \frac{p_i}{V_i},
\]
which defines \emph{observational entropy with a single coarse-graining}.\footnote{For a single coarse-graining, we can also define the coarse-grained density matrix $\R_{\mathrm{cg}}=\sum_i p_i \frac{\P_i}{V_i}$, and define $S_{\C}\equiv S_{\mathrm{vN}}(\R_{\mathrm{cg}})$. This type of definition is common in literature~\cite{wehrl1978general,almheiri2020entropy}. However, for multiple coarse-grainings that do not commute, writing observational entropy like this is not possible.}

We can view this definition as a contribution of two separate terms,
\[\label{eq:division}
S_{\C}=\mean{-\ln p_i}_{p_i}+\mean{\ln V_i}_{p_i}
=S_{\mathrm{Sh}}(p_i)+\mean{S_{\mathrm B}(i)}_{p_i},
\]
the first being the Shannon entropy of an observable,\footnote{Entropy of an observable $S_\C^{\mathrm{O}}\equiv S_{\mathrm{Sh}}(p_i)$ is sometimes also called entropy of partition~~\cite{daniel1984entropy,jost2006dynamical,anza2017information,lent2019quantum}.} describing uncertainty in obtaining a specific (macro-) measurement outcome, and the second being the mean Boltzmann entropy describing expected uncertainty regarding the (micro-)state of the system after the measurement. This entropy can be therefore interpreted as the uncertainty associated with the system if the observer were to make the measurement, without actually doing so. In other words, it can be viewed as the average uncertainty inferred about the initial (pre-measurement) state by making the measurement. If the measurement {\em were} performed, the entropy associated with the system after obtaining a measurement result $i$ would be the Boltzmann entropy $S_{\mathrm B}(i)=\ln V_i$, while the average information about this post-measurement state would be the mean value of Boltzmann entropy $\mean{S_{\mathrm B}(i)}_{p_i}$.

Unlike in classical systems, in quantum systems microstates $\ket{\psi}$ can span multiple macrostates. When associating entropy to such a state, averaging over Boltzmann entropies is therefore necessary -- the only alternative being to say that the state is in a superposition of states with distinct Boltzmann entropies~\cite{goldstein2019gibbs}. It is important to note, however, that both parts $S_{\mathrm{Sh}}(p_i)$ and  $\mean{S_{\mathrm B}(i)}_{p_i}$ are important since either of those would suffer of some pathological behavior if it was just by itself. For example, if there are $M$ measurement outcomes, the Shannon entropy is bounded by $S_{\mathrm{Sh}}(p_i)\leq \ln M$. Then, if coarse-graining is defined by a complete set of observables, then each element has dimension 1 (corresponding to rank-1 projectors), and mean Boltzmann entropy $\mean{S_{\mathrm B}(i)}_{p_i}$ is always zero, independent of the state. In either case, $S_{\mathrm{Sh}}(p_i)$ and $\mean{S_{\mathrm B}(i)}_{p_i}$ by themselves are more informative of the measurement rather than of the state of the system. It is only the sum of the two that can be interpreted as an entropy of associated with the system.

With this in mind, Observational entropy can be seen as a quantum generalization of both the Shannon and Boltzmann entropies of a measurement, with each representing a particular limit.

\section{multiple coarse-grainings}

Naturally, one can ask what entropy to attribute to a system if an observer performs not just one, but two or more measurements. This calls for a generalization of observational entropy to multiple coarse-grainings. %Commuting coarse-grainings can combined into one unifying coarse-graining (called a \emph{joint coarse-graining}), and then
Since in quantum physics two measurements do not necessarily commute, in general they cannot be combined into a single unifying coarse-graining (called a \emph{joint coarse-graining}). Joint coarse-graining exists only for commuting coarse-grainings, which means there is no obviously unique way of generalizing definition~\eqref{eq:def1}.

As shown in~\cite{safranek2019long}, however, a viable and natural option that leads to the desired properties explicated below is
\[\label{eq:oemultiple}
S_{\C_1,\dots,\C_n}\equiv-\sum_{\bi}p_\bi\ln \frac{p_\bi}{V_\bi},
\]
where multi-index $\bi=(i_1,\dots,i_n)$ denotes a set of macroscopic properties, $p_\bi=\tr[\P_{i_n}\cdots\P_{i_1}\R\P_{i_1}\cdots\P_{i_n}]$ is the probability of these properties being measured (in the given order), and $V_\bi=\tr[\P_{i_n}\cdots\P_{i_1}\cdots\P_{i_n}]$ denotes a joint Hilbert space volume of all systems that have properties $\bi=(i_1,\dots,i_n)$ measured in this order. We call $V_\bi$ the volume of the multi-macrostate $\bi$.\footnote{Even more general definition would involve generalized measurements (POVMs) which are defined by a trace-preserving ($\sum_i\hat{K}_i^\dag\hat{K}_i=\hat{I}$) set of Kraus operators $\C=\{\hat{K}_i\}$, which defines $p_\bi=\tr[\hat{K}_{i_n}\cdots\hat{K}_{i_1}\R\hat{K}_{i_1}^\dag\cdots\hat{K}_{i_n}^\dag]$ and $V_\bi=\tr[\hat{K}_{i_n}\cdots\hat{K}_{i_1}\hat{K}_{i_1}^\dag\cdots\hat{K}_{i_n}^\dag]$. Properties~\eqref{eq:bounds} and~\eqref{eq:nonincreasing} still hold~\cite{safranek2020observational}.}

Importantly, $p_\bi$, $V_\bi$ and $S_{\C_1,\dots,\C_n}$ all depend on the order of coarse-grainings. This illustrates that a different order of measurements uncovers different amounts of information about the measured properties.

Observational entropy satisfies the following properties:
\begin{align} 
S_{\mathrm{vN}}(\R)\leq S_{\C_1,\dots,\C_n}(\R)\leq \ln \mathrm{dim}\HS,\label{eq:bounds}\\ 
S_{\C_1,\dots,\C_n}(\R)\leq  S_{\C_1,\dots,\C_{n-1}}(\R). \label{eq:nonincreasing} 
\end{align}
The first property shows that observer's uncertainty about the system (measured by observational entropy) is at least the uncertainty inherent to the system (measured by the von Neumann entropy), and lower than the maximal possible uncertainty allowed by the size of the system. Observational entropy coincides with the von Neumann entropy if the sequence of measurement results in measuring the density matrix itself, which is the most informative measurement (specifically, $S_{\C_{\R}}(\R)=S_{\mathrm{vN}}(\R)$). Conversely, the uncertainty is maximal, $S_{\C_1,\dots,\C_n}= \ln \mathrm{dim}\HS$, if probabilities are proportional to the size of each macrostate, $p_\bi=V_\bi/\dim \HS$, which signifies uniform distribution over entire Hilbert space, at least within the observer's resolution. The second property shows that every additional measurement can only decrease observer's uncertainty.\footnote{One can think about equations~\eqref{eq:bounds} and~\eqref{eq:nonincreasing} in combination, and ask whether performing more measurements will always lead to the minimal uncertainty given by the von Neumann entropy. Closer analysis reveals that this is not always possible: performing a measurement that does not commute with the density matrix might irreversibly destroy some information. And when the state of the system is finally projected onto a pure state, the observational entropy is set---no additional coarse-graining will decrease it further. This also shows that initial measurements are more important than those performed later, because the later ones can uncover only information which has not been destroyed by those preceding them~\cite{safranek2019long}.}

For finite-dimensional systems, observational entropy is related to the Kullback-Leibler divergence (relative entropy) as
\[\label{eq:KL}
S_{\C_1,\dots,\C_n}=\ln \dim\HS-D_{\mathrm{KL}}(p_\bi||V_\bi/\dim \HS),
\]
which shows that observational entropy measures how much the outcome probabilities differ from those produced by a state that is uniform over the Hilbert space.\footnote{Note that $V_\bi/\dim \HS=\tr[\P_{i_n}\cdots\P_{i_1}\frac{\hat{I}}{\dim\HS}\P_{i_1}\cdots\P_{i_n}]$.} Relative entropy
$D_{\mathrm{KL}}(p_\bi||V_\bi/\dim \HS)$ may be viewed as the knowledge obtained about the state by making a series of measurements.

Details, proofs, and several other properties can be found in~\cite{safranek2019long}.

\section{classical Observational entropy}

A simpler---classical---version of observational entropy can be defined on any set $\Gamma$ endowed with a probability distribution $\rho$.\footnote{The triple $(\Gamma, \C, \rho)$ closely resembles the construction of probability space, where $\Gamma$ is the sample space and $\rho$ is the probability function, except that while $\C$ consists of disjoint events $P_i$, it does not satisfy the defining properties of event space because in general $\Gamma \notin \C$.} 
A coarse-graining $\C=\{P_i\}$ is a complete set ($\Gamma=\bigcup_iP_i$) of disjoint ($P_i\cap P_j=P_i\delta_{ij}$) subsets---\emph{macrostates}---of $\Gamma$.

%The  definition  (1)  of  observational  entropy  is  un-changed,  but  in  the  classical  case  the  probabilities  and volumes are defined as

The definition  of  observational entropy~\eqref{eq:def1}  is  unchanged, 
\[\label{eq:classOE}
S_{\C}^{\mathrm{class.}}\equiv-\sum_i p_i \ln \frac{p_i}{V_i},
\]
but  in  the  classical  case  the  probabilities  and volumes are defined differently. For a measurable set $\Gamma$, $p_i=\int_{P_i} \rho(\gamma)\ d\gamma$ and $V_i=\int_{P_i} d\gamma$ where we assume $\sum_{\gamma\in \Gamma} \rho_\gamma=1$ and $\int_\Gamma \rho(\gamma) d\gamma=1$ respectively. In case of a countable set $\Gamma$, this definition reduces to $p_i=\sum_{\gamma\in P_i} \rho_\gamma$, and $V_i$ is the cardinality (the number of elements) of $P_i$.

This definition can be easily generalized to multiple coarse-grainings~\cite{safranek2019classical}. It follows that equivalent properties to Eqs.~\eqref{eq:bounds} and \eqref{eq:nonincreasing} hold, where the von Neumann entropy is exchanged with the Shannon-Gibbs entropy $S_{\mathrm{vN}}\rightarrow S_{\mathrm{G}}=-\int_{\Gamma}\rho(\gamma) \ln \rho(\gamma) d\gamma$, and %$V=\int_{\Gamma}d\gamma$. %
$\ln \dim \HS \rightarrow \ln \int_{\Gamma}d\gamma$. 
There is, however, one crucial difference: classical coarse-grainings always commute, therefore %a joint coarse-graining always exists and
classical observational entropy does not depend on the order of coarse-grainings.

An example of classical observational entropy that was closely studied~\cite{safranek2019classical} is that defined on phase-space, where $\Gamma$ represents the phase-space, $\rho$ is the phase-space density, $\gamma=(\bx_1,\dots,\bx_N,\bp_1,\dots,\bp_N)$ is a point in phase-space, and the measure is normalized by a physically-motivated factor $d\gamma=\frac{1}{h^{3N}}d\bx_1\cdots d\bx_Nd\bp_1\cdots d\bp_N$ ($h$ being Planck's constant), ensuring that each quantum microstate, which is taking up a phase-space volume of $h^{3N}$, has volume $V=1$.

\section{Local coarse-grainings}
\label{sec-lcg}

In many situations, an one might want to consider only local measurements, in which a measuring device can only access part of a system, such as the number of particles or energy in a subsystem. This might be by choice, or -- as in the case of an event horizon -- by necessity.\footnote{In addition, in literature of coarse-grained free energies~\cite{espanol1997coarse,gao2017analytical,batchelor1967introduction,smith1950introduction,guggenheim1956statistical,callen1998thermodynamics,fisher1998renormalization,kardar2007statisticalparticles,kardar2007statisticalfields,ma2018modern,wilson1971renormalization}, one wants to find a free energy functional that depends on local variables (such as energy, particle density, magnetization...) and either study its dynamics, or critical behavior using methods of renormalization group. The current framework allows for rigorously defining these functionals, which seem to be equivalently described by Observational entropy with local coarse-grainings, for both classical and fully quantum systems.}

Consider a multipartite quantum system partitioned into local subsystems $AB\ldots C$, whose Hilbert space is the tensor product  $\HS=\HS_A\otimes\cdots\otimes\HS_C$. One can define a subclass of coarse-grainings, the \emph{local} (or \emph{product}) coarse-grainings. These are defined by
\[
\C_A\otimes\C_B\otimes\cdots\otimes\C_C\equiv\{\P_l^{A}\otimes\P_m^{B}\otimes\cdots\otimes\P_n^{C}\},
\]
where $\C_A= \{\P_l^A\}$ is a coarse-graining of $A$, and so on for the other subsystems.  These coarse-grainings correspond to local operators that only operate on one subsystem at a time.

Applying the definition \eqref{eq:def1} in such a coarse-graining yields the entropy\footnote{A local coarse-graining is equivalent to a sequence of coarse-grainings: defining trivial coarse-graining $\C_{\hat{I}}=\{\hat{I}\}$ which represents a situation where no measurement is performed, the definition can be written in terms of sequence of local measurements performed on different subsystems, as $S_{\C_A \otimes \ldots \otimes \C_C}=S_{\tilde{\C}_A,\dots,\tilde{\C}_C}$, where $\tilde{\C}_A=\C_A\otimes \C_{\hat{I}} \otimes \cdots \otimes \C_{\hat{I}}$, $\dots$, $\tilde{\C}_C=\C_{\hat{I}}\otimes \C_{\hat{I}} \otimes  \cdots \otimes \C_C$.}
\[
\label{eq:obsent-prod}
    S_{\C_A \otimes \ldots \otimes \C_C} = 
    -\sum_{lm \ldots n} p_{lm \ldots n} \log \left( \frac{p_{lm \ldots n}}{V_{lm \ldots n}} \right),
\]
where $p_{lm \ldots n} = \tr(\P^A_l \otimes \P^B_m \otimes \ldots \otimes \P^C_n \,\R)$ are the probabilities to find the system in each macrostate, and $V_{lm \ldots n} = \tr(\P^A_l \otimes \P^B_m \otimes \ldots \otimes \P^C_n)=V_lV_m\cdots V_n$ are the volumes of each macrostate.

This expression includes contributions from both the observation entropies of the subsystems, and also correlations between subsystems: simple algebraic manipulation shows that
\[
\label{eq:product-fomula}
    S_{\C_A \otimes \cdots \otimes \C_C}(\R) = 
    \Big( \sum_{X} S_{C_X}(\R_X) \Big)
    - I_{\C_A \otimes \cdots \otimes \C_C}(\R),
\]
where $X \in \{A,B,\ldots,C \}$ labels the subsystems, with $\rho_X$ the reduced density in each one, and
\[
    I_{\C_A \otimes \ldots \otimes \C_C}(\R) \equiv \sum_{lm \ldots n} p_{lm \ldots n} \log \left( \frac{p_{lm \ldots n}}{p^A_l p^B_m \ldots p^C_n} \right)
\]
is the total correlation of the joint measurement. The~$p^A_l \equiv \sum_{m\dots n}p_{lm\dots n}=\tr(\P^A_l\rho_A)$ and so on are marginal probabilities, and $I \geq 0$.

As an illustration, consider a bipartite system where the total energy $E$ is conserved. Measuring energy of the first subsystem and obtaining $E_A$ implies that the energy of the second subsystem must be $E_B=E-E_A$ (assuming negligible interaction Hamiltonian). Taking just $ S_{C_{E_A}}(\R_A)+S_{C_{E_B}}(\R_B)$ as the total entropy of the system (as done in~\cite{polkovnikov2011microscopic} for example) would overshoot the actual entropy, ignoring any relationship between them. The uncertainty arising from ignoring the other subsystem when measuring one of them would be accounted for twice, which is why $I_{\C_{E_A} \otimes \C_{E_B}}$ must be subtracted.

Because the total correlation is zero for independent systems, Eq.~\eqref{eq:product-fomula} implies that the observational entropy with local coarse-grainings is additive:
\[\label{eq:additive}
S_{\C_A \otimes \cdots \otimes \C_C}(\R_A\otimes\cdots\otimes\R_C) = 
\sum_{X} S_{C_X}(\R_X).
\]

Considering only local coarse-grainings, the lower bound on the observational entropy may be higher than the von Neumann entropy (Eq.~\eqref{eq:bounds}). Defining this entropy gap between the best possible local and the best possible global coarse-graining as
\begin{equation}
\label{eqn:ee}
    S^{\textsc{qc}}_{AB \ldots C}(\R) \equiv 
    \inf_{\C = \C_A \otimes \ldots \otimes \C_C} \Big(S_{\C}(\R) \Big) - S_{\mathrm{vN}}(\R)
\end{equation}
and studying its properties shows that this is a natural generalization of entanglement entropy to mixed and multipartite states. It reduces to the standard definition for pure bipartite states, and can be interpreted both as a measure of non-classical correlations~\cite{schindler2020entanglement}. Then, directly from the definition we obtain a very compelling bound,
\[\label{eq:local_bound}
    S_{\C_A \otimes \ldots \otimes \C_C}(\R) \geq S_{\mathrm{vN}}(\R) + S^{\textsc{qc}}_{AB \ldots C}(\R).
\]
This illustrates that any observer who can make only local measurements observers at least as much uncertainty as the inherent uncertainty in the joint state (the von Neumann entropy) plus an additional contribution (the quantum correlation entropy---``quarrelation entropy'' for short) due to their inability to make a nonlocal joint measurement.

\section{Physical applications}
\label{sec-phys}

\begin{figure}[t]
\begin{center}
\includegraphics[width=1\hsize]{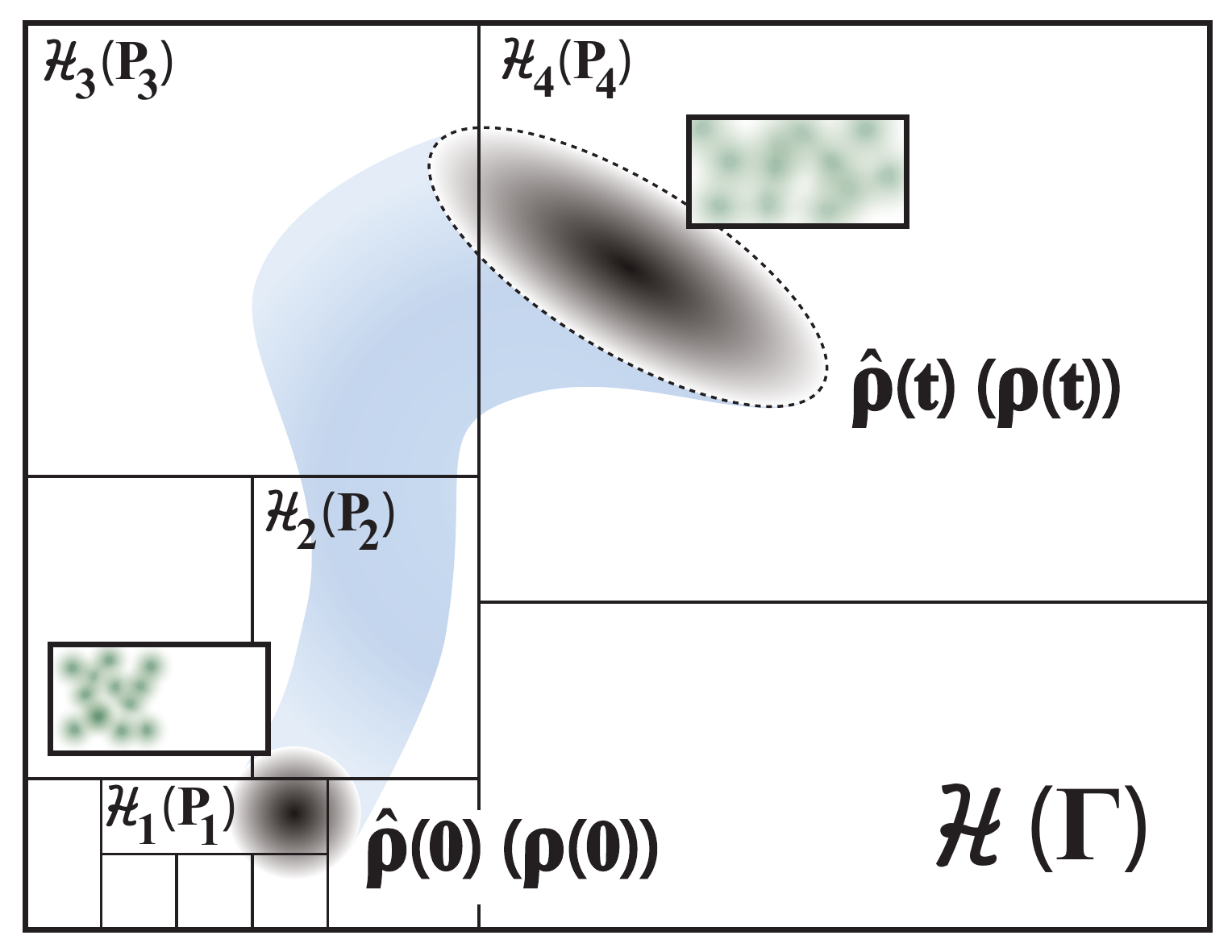}\\
\caption
{
Schematic picture of an evolution of a system through Hilbert space (phase-space), described by a density matrix $\R$ (phase-space density $\rho$), for a situation such as an expanding gas. In both quantum and classical space the density matrix (phase-space density) can span over several macrostates $\HS_i$ ($P_i$) at the same time. (Although microstates---wavefunctions in the quantum case, points in phase-space in the classical case---can span over several macrostates only in the quantum case.) As the gas expands, density matrix naturally wanders from a few small macrostates into several large macrostates, leading to an increase in observational entropy.}
\label{Fig:evolutioninphasespace}
\end{center}
\end{figure}

Because observational entropy is a quantum generalization of Boltzmann entropy, in typical situations, and especially in isolated systems, this entropy will increase with only rare downward fluctuations.\footnote{Observational entropy is therefore quite unlike the von Neumann entropy, which remains constant in an isolated system.
See Ref.~\cite{faiez2020typical} for a detailed study of fluctuations in one type of observational entropy.}
This is because the state of the system naturally evolves into the largest macrostate and/or spreads over several or many macrostates (see Fig.~\ref{Fig:evolutioninphasespace}). However, it is not clear which, if any, coarse-grainings have direct relevance to thermodynamics, which relates entropy to extensive and intensive variables such as energy, volume, particle number and temperature.
Reviewing the results of~\cite{vonNeumann1929,safranek2019short,safranek2019long,safranek2019classical}, in this section we demonstrate two coarse-grainings under which observational entropy could be considered as a definition of both equilibrium and non-equilibrium thermodynamic entropy.

The standard definition of equilibrium ``microcanonical'' entropy\footnote{Also known as the surface entropy, or the Boltzmann entropy, although in our framework, since we consider general coarse-grainings, this would be called the Boltzmann entropy with energy coarse-graining. See~ for an alternative definition of microcanonical entropy---the volume entropy---and references therein.} defines a value that depends solely on the externally measured parameters of energy, particle number, and volume. For a fixed number of particles $n$ occupying the spatial volume $\mathcal{V}$, this value is given by the energy density of states $\rho(E)$ as
\[\label{eq:microentropy}
S_{\mathrm{micro}}(E,\mathcal{V},n)=\ln\big(\rho(E)\DE\big),
\]
where $\DE$ is width of an energy shell (and experimentally is given by the resolution of the measuring apparatus measuring energy E), and $\rho(E)\DE$ is the number of states within the energy shell. Given the current framework, this equilibrium %value of 
entropy can be generalized to systems with variable energy and variable number of particles as
\[\label{eq:eqilibrium}
S_{\th}(\R)\equiv S_{\C_{\hat{N}},\C_E}(\R),
\]
where $\C_{\hat{N}}$ is the coarse-graining in the global particle number, and $\C_E$ in the global energy.\footnote{The particle coarse-graining is defined as $\C_{\hat{N}}=\{\P_n\}$, where $\P_n$ is a projector onto subspace of $n$ particles, and energy coarse-graining as $\C_E =\{\P_E\}$, where $\P_E=\sum_{E\leq \tilde E <E+\DE}\pro{\tilde E}{\tilde E}$ is a projector onto subspace of wave functions within an energy shell $[E,E+\DE)$.}
Depending on the density matrix, this formula gives microcanonical, canonical, and grand-canonical entropy, obtained when inserted with a density matrix representing each ensemble, but it can be applied to any density matrix. The microcanonical entropy is obtained, for example, when inserted with a common eigenstate of both energy and particle operator $\R=\pro{n,E}{n,E}$,\footnote{As well as when inserted with a microcanonical state $\R=\frac{1}{Z}\sum_{E\leq{\tilde E}<E+\DE}\pro{\tilde E}{\tilde E}$. The volume microcanonical entropy~ is obtained by inserting $\R=\frac{1}{Z}\sum_{0\leq{\tilde E}<E}\pro{\tilde E}{\tilde E}$, the canonical by $\R=\frac{1}{Z}e^{-\beta \hat{H}}$, and grandcanonical by $\R=\frac{1}{Z}e^{-\beta (\hat{H}-\mu \hat{N})}$.} 
\[\label{eq:microensemble}
S_{\th}(\ket{n,E})=S_{\mathrm{micro}}(E,\mathcal{V},n).
\]
For an isolated system that conserves the total number of particles, the distributions in both $E$ and $n$ stay constant,\footnote{The spatial volume $\mathcal{V}$ is assumed to be fixed implicitly here, but in general it might not be, for example when considering a work-extraction protocol using a piston.} and value of equilibrium entropy~\eqref{eq:eqilibrium} remains constant in time as expected.\footnote{Eq.~\eqref{eq:eqilibrium} can be generalized to any number of conserved observables, see Eq.~\eqref{eq:generaleqentropy}} Note that while the general definition of observational entropy depends on the choice of a coarse-graining, the coarse-graining in the notion of thermodynamic entropy introduced here depends primarily on the Hamiltonian and other conserved quantities, and therefore is given primarily by the system and not by the observer.

\begin{figure}[htbp]
\begin{center}
\includegraphics[width=1\hsize]{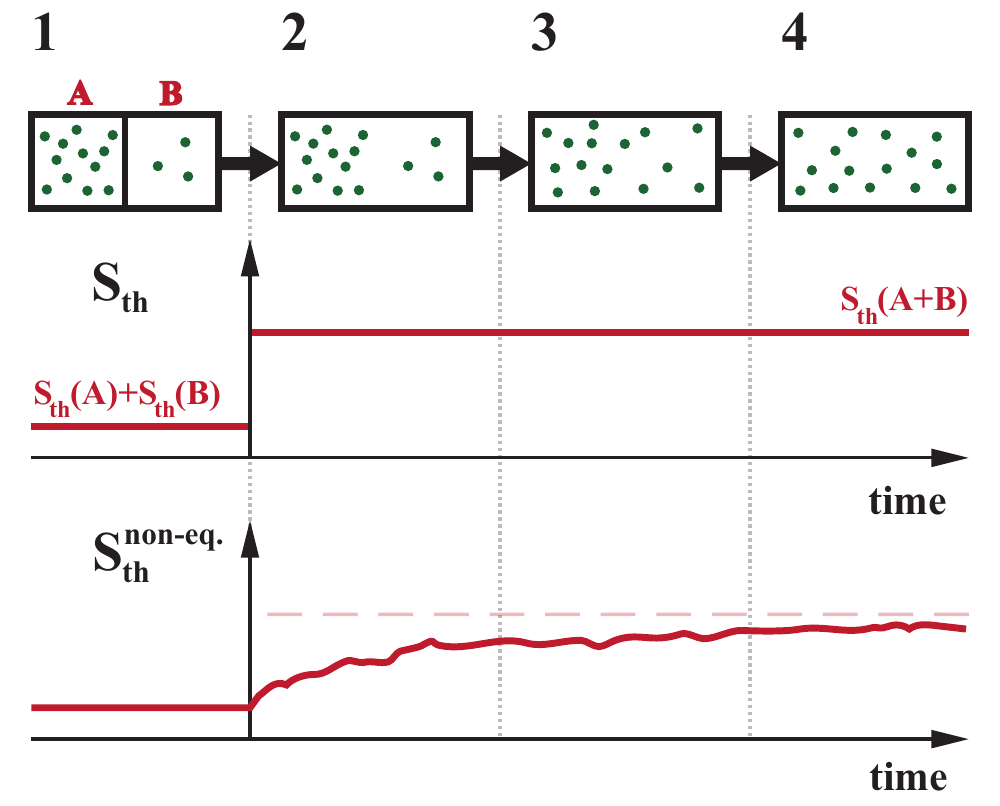}
\caption
{
Equilibrium thermodynamic entropy $S_{\th}$, calculated either from Eq.~\eqref{eq:microentropy} or from its generalization~\eqref{eq:eqilibrium}, discontinuously increases from $1 \rightarrow 2$, because the Hamiltonian (or equivalently, boundary conditions) discontinuously changes. Then it stays constant. Non-equilibrium thermodynamic entropy describes the \emph{dynamical} process of equilibration, i.e., a measure that depends on the state of the system rather than on the boundary conditions. Such measure is expected to stay constant as $1 \rightarrow 2$, to increase during $2 \rightarrow 4$, and to be approximately equal to equilibrium thermodynamic entropy at points $1$ and $4$, when the system is in equilibrium.
}
\label{Fig:expansionscheme}
\end{center}
\end{figure}

\begin{figure}[htbp]
\begin{center}
\includegraphics[width=0.8\hsize]{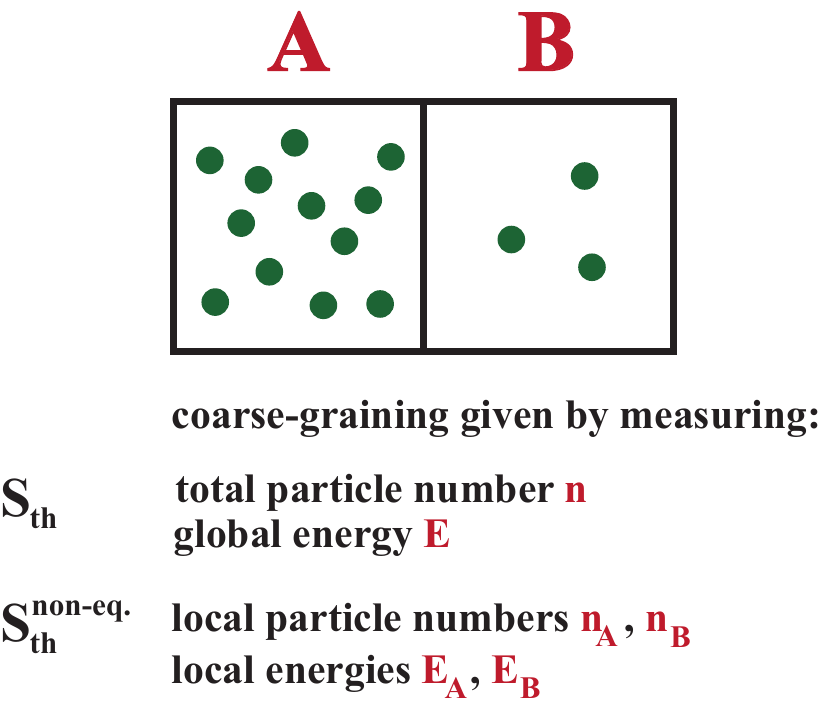}
\caption
{
Equilibrium thermodynamic entropy $S_{\th}$ is given by observational entropy coarse-grained in global observables that are conserved during the time evolution (such as the global energy and the total particle number), while non-equilibrium thermodynamic entropy is given by coarse-graining in local observables (such as local energies and local particle numbers). Assuming a weakly interacting system, if local evolution of the system is such that the density matrix (phase-space density) explores uniformly each shell of the global observables during its time evolution (for example due to ergodicity), the non-equilibrium thermodynamic entropy will converge to the equilibrium thermodynamic entropy in the long-time limit.
}
\label{Fig:cg}
\end{center}
\end{figure}

In contrast to equilibrium entropy, non-equilibrium thermodynamic entropy should not depend on the globally measured parameters of the system, but on the parameters measured locally, which brings in dependence on the non-equilibrium features of the state of the system. Additionally, it should correspond to the equilibrium value when the system is in equilibrium (see Fig.~\ref{Fig:expansionscheme}). Assuming that an experimentalist can measure energies and particle numbers locally, the non-equilibrium thermodynamic entropy is defined as\footnote{Here we assume multipartite system $\HS=\HS_1\otimes\cdots\HS_m$. Each $\HS_i$ is a space of all quantum states that can occur within a spatial region of volume $\mathcal{V}_i$, and $\C_{\hat{N}_i}$ and $\C_{E_i}$ correspond to a particle and energy measurement of this spatial region.}~\cite{safranek2019long,safranek2019classical}
\[\label{eq:noneqth}
S_{\th}^{\mathrm{non-eq.}}(\R)\equiv S_{\C_{\hat{N}_1}\!\!\otimes\cdots\otimes \C_{\hat{N}_m}\!,\  \C_{E_1}\!\!\otimes\cdots\otimes \C_{E_m}}(\R).
\]
(See Fig.~\ref{Fig:cg}.) Per Eq.~\eqref{eq:additive}, this entropy is additive on independent systems. For the special case of common eigenstates of local particle and energy operators, this entropy gives the sum of local microcanonical entropies:
\[
S_{\th}^{\mathrm{non-eq.}}(\ket{n_1,E_1}\otimes\cdots\otimes\ket{n_m,E_m})=\sum_{i=1}^m S_{\mathrm{micro}}(E_i,\mathcal{V}_i,n_i).
\]
In the long-time limit, assuming weakly interacting (so that interaction energy between subsystems is negligible) particle conserving (so that $S_\th$ is constant) non-integrable (so it thermalizes well, because there are not infinitely many conserved quantities) Hamiltonian, non-equilibrium entropy converges to the equilibrium value, 
\[
S_{\th}^{\mathrm{non-eq.}}\xrightarrow{t\rightarrow \infty} S_{\th},
\]
This convergence holds up to some finite size and finite particle number effects (see the Appendix and Refs.~\cite{safranek2019long,safranek2019classical} for details).

The limit acts as its approximate upper bound\footnote{Proven to be an approximate upper bound in weakly interacting systems~\cite{safranek2019long}.}, which together with Eq.~\eqref{eq:local_bound} gives\footnote{Additional exact bounds are given by Eqs.~\eqref{eq:bounds},~\eqref{eq:nonincreasing}. When local Hamiltonians and local particle operators commute, $[\hat{H}_i,\hat{N}_i]=0$, Eq.~\eqref{eq:nonincreasing} provides two bounds: one connected to observational entropy with just local particle numbers, and one with local energies.}
\[
  S_{\mathrm{vN}} + S^{\textsc{qc}}_{12 \ldots m} \leq S_{\th}^{\mathrm{non-eq.}} \lesssim S_{\th}
\]

\begin{table*}[t!]
\caption{\label{tab:relationship}  Relationship of observational entropy with previously defined entropies (no color) and those defined by us (red). Several quantities are known by different names; we include all of these names here. Type of entropy refers to: C (coarse-graining based entropy), I (information-theoretic), T (thermodynamic).}
\begin{tabular}{|l|l|l|l|l|}
\hline
\rowcolor{blue!10} Entropy & Type & Definition & Relationship & Details \\
\hline
Boltzmann entropy & C/T & $S_\C^{\mathrm{B}}\equiv\ln V_i$ & special case: for $p_i=1$, $S_\C^{\mathrm{B}}=S_\C$  &    \\
Coarse-grained entropy & C & $S_\C^{\mathrm{cg}}\equiv-\sum_i p_i\ln \frac{p_i}{V_i}$  & single coarse-graining: $S_\C^{\mathrm{cg}}=S_\C$ & Eq.~\eqref{eq:def1}   \\
\rowcolor{red!10} Observational entropy & C & $S_{\C_1,\dots,\C_n}\equiv-\sum_{\bi}p_\bi\ln \frac{p_\bi}{V_\bi}$ &  &  Eq.~\eqref{eq:oemultiple}  \\

Entropy of an observable/entropy & C & $S_\C^{\mathrm{O}}\equiv-\sum_i p_i\ln p_i$ & equal to $S_{\C}$ when all $V_i=1$ & Eq.~\eqref{eq:division}   \\
of partition &  &  &  &    \\
Shannon entropy & I & $S_{\mathrm{Sh}}(\{p_j\}_j)\equiv-\sum_j p_j\ln p_j$ & $S_{\C}=S_{\mathrm{Sh}}(\{\frac{p_i}{V_i}\}_{i,k})$  & Eq.~\eqref{eq:shannon}   \\
Relative entropy & I & $D_{\mathrm{KL}}(p_\bi||q_\bi)\equiv\sum_\bi p_\bi\ln\frac{p_\bi}{q_\bi}$ & 
$S_{\C_1,\dots,\C_n}\!\!=\!\ln \dim\HS\!-\!D_{\mathrm{KL}}(p_\bi||\tfrac{V_\bi}{\dim \HS})$ &  Eq.~\eqref{eq:KL}  \\
Gibbs/Differential entropy & T/I & $S_{\mathrm{G}}\equiv-\int_{\Gamma}\rho(\gamma) \ln \rho(\gamma) d\gamma$ & $S_{\C_1,\dots,\C_n}^{\mathrm{class.}}\geq S_{\mathrm{G}}$  &  Eq.~\eqref{eq:classOE}  \\
Von Neumann entropy & I & $S_{\mathrm{vN}}\equiv-\tr[\R\ln\R]$ & $S_{\C_1,\dots,\C_n}\geq S_{\mathrm{vN}}$, %$S_\C_{\R}=S_{\mathrm{vN}}$,
\ $S_\C=S_{\mathrm{vN}}\big(\sum_ip_i\frac{\P_i}{V_i}\big)$  &  Eq.~\eqref{eq:bounds}  \\
Entanglement entropy & I & $S^{\mathrm{ent}}(\ket{\psi})\equiv S_{\mathrm{vN}}(\R_A)=S_{\mathrm{vN}}(\R_B)$ & $S^{\mathrm{ent}}(\ket{\psi})=S^{\textsc{qc}}_{AB}(\ket{\psi})$ & Eq.~\eqref{eqn:ee} \\
 &  & & $\ \ \ \ \ \ \ \ \ \ \ \ \ \! =\displaystyle{\inf_{\C_A\otimes\C_B}}S_{\C_A\otimes\C_B}(\ket{\psi})$ &    \\
\rowcolor{red!10} Equilibrium thermodynamic entropy & T & $S_{\th}\equiv S_{\C_{\hat{N}},\C_E}$ & & Eq.~\eqref{eq:eqilibrium}  \\
Microcanonical, canonical, & T & $S_{\mathrm{ensemble}}$%\equiv\ln (\rho(E)\Delta E),$  
& 
$S_{\mathrm{ensemble}}=S_{\th}(\R_{\mathrm{ensemble}})$ 
  &  Eq.~\eqref{eq:eqilibrium}  \\
grandcanonical entropy  &  & for example $S_{\mathrm{micro}}(E,\mathcal{V},n)$%$\ln {Z_\beta}\!+\!\beta\mean{E}, \ln {Z_{\beta,\mu}}\!+\!\beta\mean{E}\!+\!\mu\mean{N}$ 
& $S_{\mathrm{micro}}(E,\mathcal{V},n)=S_{\th}(\R_{\mathrm{micro}})$   &  Eq.~\eqref{eq:microensemble}  \\
\rowcolor{red!10} Non-equilibrium thermodynamic & T & 
$S_{\th}^{\mathrm{non-eq.}}\!\equiv\! S_{\C_{\hat{N}_1}\!\!\otimes\cdots\otimes \C_{\hat{N}_m}\!,\C_{E_1}\!\!\otimes\cdots\otimes \C_{E_m}}$ &  %$S_{\th} \gtrsim S_{\th}^{\mathrm{non-eq.}} \geq S_{\mathrm{vN}} + S^{\mathrm{ent}}_{1\ldots m}$,
$S_{\th}^{\mathrm{non-eq.}}\xrightarrow{t\rightarrow \infty} S_{\th}$, \ $S_{\th}^{\mathrm{non-eq.}}\lesssim S_{\th}$ & Eq.~\eqref{eq:noneqth} \\
\rowcolor{red!10} entropy &  & &
& \\
\rowcolor{red!10} Quantum correlation entropy/ & I & 
    $\displaystyle{S^{\textsc{qc}}_{AB \ldots C} \!\equiv\! 
    \inf_{\C = \C_A \otimes \ldots \otimes \C_C} \!\big(S_{\C} \big) \!-\! S_{\mathrm{vN}}}$ & 
    $S_{\C_A \otimes \ldots \otimes \C_C} \geq S_{\mathrm{vN}} + S^{\textsc{qc}}_{AB \ldots C}$ & Eq.~\eqref{eqn:ee}  \\
\rowcolor{red!10} relative entropy of quantum discord/ &  & 
     &  &
      \\
\rowcolor{red!10} zero-way quantum deficit &  & 
     &  &
      \\
\hline
\end{tabular}
\end{table*}

The non-equilibrium thermodynamic entropy therefore describes the dynamical process of thermalization: starting as the sum of local entropies of independent subsystems, and after these subsystems start to interact, it grows to the global equilibrium entropy as the system thermalizes. It also has a very intuitive and compelling operational interpretation: At some intermediate time~$t$ (when the system has only partially equilibrated) its value can be interpreted as the equilibrium thermodynamic entropy the system would attain in the long-time limit if (hypothetically) starting from time $t$ the subsystems were not allowed to exchange either energy or particles~\cite{safranek2019classical}.

In non-equilibrium thermodynamic entropy, the choice of coarse-graining depends partially on the observer. It is the observer who chooses the partitioning into smaller subsystems, and then the coarse-graining is given by Hamiltonians of the subsystems themselves. The observer's choice, however, would be naturally informed: it makes more sense to separate a joint system of a melting ice cube in a cup of water into a subsystem of ice and a subsystem of water. While other partitions are equally valid,\footnote{For example the first half of ice and the first half of water would together form the first subsystem, while the rest would form the second subsystem.} they might not properly describe the process of a melting ice: the initial value of entropy could be quite large (close to the limit imposed by the equilibrium entropy), thus growing very little before achieving its maximum.

Finally, let us move into complete generality. In Eq.~\eqref{eq:eqilibrium}, the particle and energy coarse-graining has been chosen because it is standard for systems to conserve particle numbers, and this definition corresponds to the standard notions of equilibrium entropy. In general, there can be any number of conserved quantities (observables) $\hat{A}_1, \hat{A_2},\dots$ in the system (typically one of them being energy). Having these conserved quantities, we define equilibrium entropy as
\[\label{eq:generaleqentropy}
S^{\mathrm{eq.}}=S_{\C_{\hat{A}_1},\ \C_{\hat{A}_2},}\dots
\]
By definition, this expression stays constant in time. The non-equilibrium entropy can be then defined using local versions of these observables, as
\[
S^{\mathrm{non-eq.}}\equiv S_{\C_{\hat{A}_{11}}\!\!\otimes\cdots\otimes \C_{\hat{A}_{1m}}\!,\  \C_{\hat A_{21}}\!\!\otimes\cdots\otimes \C_{\hat A_{2m}},}\dots
\]
It is expected that in weakly interacting isolated systems, the non-equlibirum entropy will grow to the equilibrium entropy as the system thermalizes. Interpretation follows the same pattern: non-equilibrium entropy is the value of entropy the isolated system would achieve in the long-time limit, if the subsystems were not allowed to exchange any of the values of the observables. For example, observable $\hat{A}_1$ cannot flow from subsystem $1$ to subsystem $m$, meaning that the mean values (and the probability distributions) of $\hat{A}_{11}$ and $\hat{A}_{1m}$ remain constant.

There are several other coarse-grainings and corresponding observational entropies that could be considered, each with a different intepretation. For those, please see the Appendix, which also contains a few technical details and references.

%other entropies in the appendix

\section{Summary and Discussion}
\label{sec-conc}

As discussed in the introduction, there are many types of entropy in physics; some entropies arise from information theoretic concepts, some from a type of coarse-graining, and some from a thermodynamic perspective. 

Each of these entropies have a different purpose. Generally, an entropy can be considered a measure of missing information regarding the particular state of a system; different entropies measure different sources of such uncertainty. Given a closed system, lack of knowledge of the state of a system, the inability to distinguish different states of a system, and a restriction to measuring a subsystem of the system lead to, respectively, Gibbs/von Neumann, Boltzmann, and (in quantum systems) entanglement entropies. In turn, these relate in some sense to the {\em thermodynamic} entropy that is maximized in equilibrium, relates work/energy to temperature, and tends to rise in closed systems.

Observational entropy is primarily a coarse-graining based entropy which generalizes Boltzmann entropy to quantum systems. It depends on a density function or operator $\R$, a (vector of) coarse-graining(s) ${\boldsymbol{\C}} \equiv (\C_1,\dots,\C_n)$, and (potentially) a partitioning of the system into local subsystems. It is interpreted as the amount of information an observer would infer about the initial state of the system, if he or she were to perform a measurement in a basis given by the coarse-graining.

As summarized in this paper, the different senses of entropy outlined above can be both conceptually and mathematically unified in the framework of observational entropy. In particular, observational entropy transforms into other definitions of entropy given a special choice of coarse-graining or when minimizing over different types of coarse-grainings. This formalism allows for inclusion of multiple and even non-commuting coarse-grainings, and can be further generalized to POVMs and likely to other frameworks for describing acquisition of knowledge.

Some key results in quantum systems are:
\begin{itemize}
    \item With a ``fine'' graining into individual states in an appropriate basis, $S_{\boldsymbol{\C}}$ yields the von Neumann entropy.
\item For \emph{general} coarse-grainings, $S_{\boldsymbol{\C}}$ is a form of Boltzmann entropy that is bounded below by the von Neumann entropy. 
\item For \emph{local} coarse-grainings, $S_{\boldsymbol{\C}}$ is bounded by the sum of the von Neumann and the ``quantum correlation entropy,'' which generalizes entanglement entropy.
\item The \emph{equilibrium} thermodynamic entropy is given by coarse-grainings in \emph{global} energy and \emph{global} particle number (or by coarse-graining in other globally conserved quantities), and it generalizes other equilibrium entropies (such as microcanonical, canonical, and grand-canonical entropy). \item \emph{Non-equilibrium} thermodynamic entropy is given by coarse-graining in \emph{local} energy and \emph{local} particle numbers, is additive on independent systems, and equal to the equilibrium entropy when the system is in equilibrium. 
\item Suitably applied to open systems, $S_{\boldsymbol{\C}}$ reduces to the standard formalism under the assumption of an infinite thermal bath.
\end{itemize}
See Table~\ref{tab:relationship} for more relations between observational entropy and other quantities.

Various forms of the second law of thermodynamics emerge in observational entropy as they do in the more special special cases.  Boltzmann entropy tends to rise (while occasionally fluctuating down) due to wandering of a system into higher-entropy macrostates; entanglement entropy tends to rise due to the forging of entanglement between two interacting subsystems; von Neumann entropy is forbidden from decreasing (or increasing) via information preservation in unitary dynamics; the total thermodynamic entropy tends to increase due to heat flowing from a warmer to a colder body.  All of these effects are reflected in the dynamics of observational entropy.

Given the generality of this framework,
we expect it will have many applications where its well-defined conceptual and mathematical underpinnings could bring clarity -- for example in efficient discussion of the Gibbs paradox, in fluctuation theorems for both isolated and open quantum systems, in studying differences between thermalization of classical and quantum systems, in generalization of renormalization group methods to many-body systems in which quantum effects are important, and in systems with strong or long-range interactions (including gravity.)

Most broadly, it may be particularly useful in contexts in which the second thermodynamic law is used as a fundamental constraint on a total entropy that is a sum of different versions of entropy.  This includes for example Maxwell's demon and Szilard's engine (and information engines in general, where Shannon and thermodynamic entropy are combined), the ``generalized second laws'' (where horizon and statistical entropy are combined), and in cosmology (where all manner of entropies are summed and assumed to increase).  We hope that the observational entropy framework, which can accommodate many types, can be used to give more crisp and explicit mathematical and conceptual meaning to such arguments.

\begin{acknowledgments}
This research was supported by the Foundational Questions Institute (FQXi.org), of which AA is Associate Director, and by the Faggin Presidential Chair Fund. D\v S acknowledges additional funding by the Institute
for Basic Science in Korea (IBS-R024-Y2
 and IBS-R024-D1).
\end{acknowledgments}

\bibliography{biblio}

\begin{thebibliography}{43}%
\makeatletter
\providecommand \@ifxundefined [1]{%
 \@ifx{#1\undefined}
}%
\providecommand \@ifnum [1]{%
 \ifnum #1\expandafter \@firstoftwo
 \else \expandafter \@secondoftwo
 \fi
}%
\providecommand \@ifx [1]{%
 \ifx #1\expandafter \@firstoftwo
 \else \expandafter \@secondoftwo
 \fi
}%
\providecommand \natexlab [1]{#1}%
\providecommand \enquote  [1]{``#1''}%
\providecommand \bibnamefont  [1]{#1}%
\providecommand \bibfnamefont [1]{#1}%
\providecommand \citenamefont [1]{#1}%
\providecommand \href@noop [0]{\@secondoftwo}%
\providecommand \href [0]{\begingroup \@sanitize@url \@href}%
\providecommand \@href[1]{\@@startlink{#1}\@@href}%
\providecommand \@@href[1]{\endgroup#1\@@endlink}%
\providecommand \@sanitize@url [0]{\catcode `\\12\catcode `\$12\catcode
  `\&12\catcode `\#12\catcode `\^12\catcode `\_12\catcode `\%12\relax}%
\providecommand \@@startlink[1]{}%
\providecommand \@@endlink[0]{}%
\providecommand \url  [0]{\begingroup\@sanitize@url \@url }%
\providecommand \@url [1]{\endgroup\@href {#1}{\urlprefix }}%
\providecommand \urlprefix  [0]{URL }%
\providecommand \Eprint [0]{\href }%
\providecommand \doibase [0]{http://dx.doi.org/}%
\providecommand \selectlanguage [0]{\@gobble}%
\providecommand \bibinfo  [0]{\@secondoftwo}%
\providecommand \bibfield  [0]{\@secondoftwo}%
\providecommand \translation [1]{[#1]}%
\providecommand \BibitemOpen [0]{}%
\providecommand \bibitemStop [0]{}%
\providecommand \bibitemNoStop [0]{.\EOS\space}%
\providecommand \EOS [0]{\spacefactor3000\relax}%
\providecommand \BibitemShut  [1]{\csname bibitem#1\endcsname}%
\let\auto@bib@innerbib\@empty
%</preamble>
\bibitem [{\citenamefont {{{\v{S}}afr{\'a}nek}}\ \emph
  {et~al.}(2019{\natexlab{a}})\citenamefont {{{\v{S}}afr{\'a}nek}},
  \citenamefont {{Deutsch}},\ and\ \citenamefont
  {{Aguirre}}}]{safranek2019short}%
  \BibitemOpen
  \bibfield  {author} {\bibinfo {author} {\bibfnamefont {Dominik}\ \bibnamefont
  {{{\v{S}}afr{\'a}nek}}}, \bibinfo {author} {\bibfnamefont {J.~M.}\
  \bibnamefont {{Deutsch}}}, \ and\ \bibinfo {author} {\bibfnamefont {Anthony}\
  \bibnamefont {{Aguirre}}},\ }\bibfield  {title} {\enquote {\bibinfo {title}
  {{Quantum coarse-grained entropy and thermodynamics}},}\ }\href {\doibase
  10.1103/PhysRevA.99.010101} {\bibfield  {journal} {\bibinfo  {journal}
  {\pra}\ }\textbf {\bibinfo {volume} {99}},\ \bibinfo {eid} {010101} (\bibinfo
  {year} {2019}{\natexlab{a}})},\ \Eprint {http://arxiv.org/abs/1707.09722}
  {arXiv:1707.09722 [quant-ph]} \BibitemShut {NoStop}%
\bibitem [{\citenamefont {{{\v{S}}afr{\'a}nek}}\ \emph
  {et~al.}(2019{\natexlab{b}})\citenamefont {{{\v{S}}afr{\'a}nek}},
  \citenamefont {{Deutsch}},\ and\ \citenamefont
  {{Aguirre}}}]{safranek2019long}%
  \BibitemOpen
  \bibfield  {author} {\bibinfo {author} {\bibfnamefont {Dominik}\ \bibnamefont
  {{{\v{S}}afr{\'a}nek}}}, \bibinfo {author} {\bibfnamefont {J.~M.}\
  \bibnamefont {{Deutsch}}}, \ and\ \bibinfo {author} {\bibfnamefont {Anthony}\
  \bibnamefont {{Aguirre}}},\ }\bibfield  {title} {\enquote {\bibinfo {title}
  {{Quantum coarse-grained entropy and thermalization in closed systems}},}\
  }\href {\doibase 10.1103/PhysRevA.99.012103} {\bibfield  {journal} {\bibinfo
  {journal} {\pra}\ }\textbf {\bibinfo {volume} {99}},\ \bibinfo {eid} {012103}
  (\bibinfo {year} {2019}{\natexlab{b}})},\ \Eprint
  {http://arxiv.org/abs/1803.00665} {arXiv:1803.00665 [quant-ph]} \BibitemShut
  {NoStop}%
\bibitem [{\citenamefont {{{\v{S}}afr{\'a}nek}}\ \emph
  {et~al.}(2019{\natexlab{c}})\citenamefont {{{\v{S}}afr{\'a}nek}},
  \citenamefont {{Aguirre}},\ and\ \citenamefont
  {{Deutsch}}}]{safranek2019classical}%
  \BibitemOpen
  \bibfield  {author} {\bibinfo {author} {\bibfnamefont {Dominik}\ \bibnamefont
  {{{\v{S}}afr{\'a}nek}}}, \bibinfo {author} {\bibfnamefont {Anthony}\
  \bibnamefont {{Aguirre}}}, \ and\ \bibinfo {author} {\bibfnamefont {J.~M.}\
  \bibnamefont {{Deutsch}}},\ }\bibfield  {title} {\enquote {\bibinfo {title}
  {{Classical dynamical coarse-grained entropy and comparison with the quantum
  version}},}\ }\href@noop {} {\bibfield  {journal} {\bibinfo  {journal}
  {eprint}\ } (\bibinfo {year} {2019}{\natexlab{c}})},\ \Eprint
  {http://arxiv.org/abs/1905.03841} {arXiv:1905.03841 [cond-mat.stat-mech]}
  \BibitemShut {NoStop}%
\bibitem [{\citenamefont {{Strasberg}}(2019)}]{strasberg2019entropy}%
  \BibitemOpen
  \bibfield  {author} {\bibinfo {author} {\bibfnamefont {Philipp}\ \bibnamefont
  {{Strasberg}}},\ }\bibfield  {title} {\enquote {\bibinfo {title} {{Entropy
  production as change in observational entropy}},}\ }\href@noop {} {\bibfield
  {journal} {\bibinfo  {journal} {eprint}\ } (\bibinfo {year} {2019})},\
  \Eprint {http://arxiv.org/abs/1906.09933} {arXiv:1906.09933
  [cond-mat.stat-mech]} \BibitemShut {NoStop}%
\bibitem [{\citenamefont {{Faiez}}\ \emph {et~al.}(2020)\citenamefont
  {{Faiez}}, \citenamefont {{{\v{S}}afr{\'a}nek}}, \citenamefont {{Deutsch}},\
  and\ \citenamefont {{Aguirre}}}]{faiez2020typical}%
  \BibitemOpen
  \bibfield  {author} {\bibinfo {author} {\bibfnamefont {Dana}\ \bibnamefont
  {{Faiez}}}, \bibinfo {author} {\bibfnamefont {Dominik}\ \bibnamefont
  {{{\v{S}}afr{\'a}nek}}}, \bibinfo {author} {\bibfnamefont {J.~M.}\
  \bibnamefont {{Deutsch}}}, \ and\ \bibinfo {author} {\bibfnamefont {Anthony}\
  \bibnamefont {{Aguirre}}},\ }\bibfield  {title} {\enquote {\bibinfo {title}
  {{Typical and extreme entropies of long-lived isolated quantum systems}},}\
  }\href {\doibase 10.1103/PhysRevA.101.052101} {\bibfield  {journal} {\bibinfo
   {journal} {\pra}\ }\textbf {\bibinfo {volume} {101}},\ \bibinfo {eid}
  {052101} (\bibinfo {year} {2020})},\ \Eprint
  {http://arxiv.org/abs/1908.07083} {arXiv:1908.07083 [quant-ph]} \BibitemShut
  {NoStop}%
\bibitem [{\citenamefont {{Strasberg}}\ and\ \citenamefont
  {{Winter}}(2020)}]{strasberg2020heat}%
  \BibitemOpen
  \bibfield  {author} {\bibinfo {author} {\bibfnamefont {Philipp}\ \bibnamefont
  {{Strasberg}}}\ and\ \bibinfo {author} {\bibfnamefont {Andreas}\ \bibnamefont
  {{Winter}}},\ }\bibfield  {title} {\enquote {\bibinfo {title} {{Heat, Work
  and Entropy Production in Open Quantum Systems: A Microscopic Approach Based
  on Observational Entropy}},}\ }\href@noop {} {\bibfield  {journal} {\bibinfo
  {journal} {eprint}\ } (\bibinfo {year} {2020})},\ \Eprint
  {http://arxiv.org/abs/2002.08817} {arXiv:2002.08817 [quant-ph]} \BibitemShut
  {NoStop}%
\bibitem [{\citenamefont {{Schindler}}\ \emph {et~al.}(2020)\citenamefont
  {{Schindler}}, \citenamefont {{{\v{S}}afr{\'a}nek}},\ and\ \citenamefont
  {{Aguirre}}}]{schindler2020entanglement}%
  \BibitemOpen
  \bibfield  {author} {\bibinfo {author} {\bibfnamefont {Joseph}\ \bibnamefont
  {{Schindler}}}, \bibinfo {author} {\bibfnamefont {Dominik}\ \bibnamefont
  {{{\v{S}}afr{\'a}nek}}}, \ and\ \bibinfo {author} {\bibfnamefont {Anthony}\
  \bibnamefont {{Aguirre}}},\ }\bibfield  {title} {\enquote {\bibinfo {title}
  {{Entanglement entropy from coarse-graining in pure and mixed multipartite
  systems}},}\ }\href@noop {} {\bibfield  {journal} {\bibinfo  {journal} {arXiv
  e-prints}\ ,\ \bibinfo {eid} {arXiv:2005.05408}} (\bibinfo {year} {2020})},\
  \Eprint {http://arxiv.org/abs/2005.05408} {arXiv:2005.05408 [quant-ph]}
  \BibitemShut {NoStop}%
\bibitem [{\citenamefont {{Riera-Campeny}}\ \emph {et~al.}(2020)\citenamefont
  {{Riera-Campeny}}, \citenamefont {{Sanpera}},\ and\ \citenamefont
  {{Strasberg}}}]{riera2020quantum}%
  \BibitemOpen
  \bibfield  {author} {\bibinfo {author} {\bibfnamefont {Andreu}\ \bibnamefont
  {{Riera-Campeny}}}, \bibinfo {author} {\bibfnamefont {Anna}\ \bibnamefont
  {{Sanpera}}}, \ and\ \bibinfo {author} {\bibfnamefont {Philipp}\ \bibnamefont
  {{Strasberg}}},\ }\bibfield  {title} {\enquote {\bibinfo {title} {{Quantum
  systems correlated with a finite bath: nonequilibrium dynamics and
  thermodynamics}},}\ }\href@noop {} {\bibfield  {journal} {\bibinfo  {journal}
  {arXiv e-prints}\ ,\ \bibinfo {eid} {arXiv:2008.02184}} (\bibinfo {year}
  {2020})},\ \Eprint {http://arxiv.org/abs/2008.02184} {arXiv:2008.02184
  [quant-ph]} \BibitemShut {NoStop}%
\bibitem [{\citenamefont {von Neumann}(2010)}]{von2010proof}%
  \BibitemOpen
  \bibfield  {author} {\bibinfo {author} {\bibfnamefont {John}\ \bibnamefont
  {von Neumann}},\ }\bibfield  {title} {\enquote {\bibinfo {title} {Proof of
  the ergodic theorem and the h-theorem in quantum mechanics},}\ }\href
  {\doibase 10.1140/epjh/e2010-00008-5} {\bibfield  {journal} {\bibinfo
  {journal} {The European Physical Journal H}\ }\textbf {\bibinfo {volume}
  {35}},\ \bibinfo {pages} {201--237} (\bibinfo {year} {2010})},\ \Eprint
  {http://arxiv.org/abs/https://arxiv.org/abs/1003.2133}
  {https://arxiv.org/abs/1003.2133} \BibitemShut {NoStop}%
\bibitem [{\citenamefont {von Neumann}(1955)}]{von1955mathematical}%
  \BibitemOpen
  \bibfield  {author} {\bibinfo {author} {\bibfnamefont {John}\ \bibnamefont
  {von Neumann}},\ }\enquote {\bibinfo {title} {Mathematical foundations of
  quantum mechanics},}\ \ (\bibinfo  {publisher}
  {\href{http://press.princeton.edu/titles/2113.html}{Princeton university
  press}},\ \bibinfo {year} {1955})\ pp.\ \bibinfo {pages}
  {410--416}\BibitemShut {NoStop}%
\bibitem [{\citenamefont {{Wehrl}}(1978)}]{wehrl1978general}%
  \BibitemOpen
  \bibfield  {author} {\bibinfo {author} {\bibfnamefont {Alfred}\ \bibnamefont
  {{Wehrl}}},\ }\bibfield  {title} {\enquote {\bibinfo {title} {{General
  properties of entropy}},}\ }\href {\doibase 10.1103/RevModPhys.50.221}
  {\bibfield  {journal} {\bibinfo  {journal} {Rev. Mod. Phys.}\ }\textbf
  {\bibinfo {volume} {50}},\ \bibinfo {pages} {221--260} (\bibinfo {year}
  {1978})}\BibitemShut {NoStop}%
\bibitem [{\citenamefont {{Gemmer}}\ and\ \citenamefont
  {{Steinigeweg}}(2014)}]{gemmer2014entropy}%
  \BibitemOpen
  \bibfield  {author} {\bibinfo {author} {\bibfnamefont {Jochen}\ \bibnamefont
  {{Gemmer}}}\ and\ \bibinfo {author} {\bibfnamefont {Robin}\ \bibnamefont
  {{Steinigeweg}}},\ }\bibfield  {title} {\enquote {\bibinfo {title} {{Entropy
  increase in K-step Markovian and consistent dynamics of closed quantum
  systems}},}\ }\href {\doibase 10.1103/PhysRevE.89.042113} {\bibfield
  {journal} {\bibinfo  {journal} {\pre}\ }\textbf {\bibinfo {volume} {89}},\
  \bibinfo {eid} {042113} (\bibinfo {year} {2014})}\BibitemShut {NoStop}%
\bibitem [{\citenamefont {{Almheiri}}\ \emph {et~al.}(2020)\citenamefont
  {{Almheiri}}, \citenamefont {{Hartman}}, \citenamefont {{Maldacena}},
  \citenamefont {{Shaghoulian}},\ and\ \citenamefont
  {{Tajdini}}}]{almheiri2020entropy}%
  \BibitemOpen
  \bibfield  {author} {\bibinfo {author} {\bibfnamefont {Ahmed}\ \bibnamefont
  {{Almheiri}}}, \bibinfo {author} {\bibfnamefont {Thomas}\ \bibnamefont
  {{Hartman}}}, \bibinfo {author} {\bibfnamefont {Juan}\ \bibnamefont
  {{Maldacena}}}, \bibinfo {author} {\bibfnamefont {Edgar}\ \bibnamefont
  {{Shaghoulian}}}, \ and\ \bibinfo {author} {\bibfnamefont {Amirhossein}\
  \bibnamefont {{Tajdini}}},\ }\bibfield  {title} {\enquote {\bibinfo {title}
  {{The entropy of Hawking radiation}},}\ }\href@noop {} {\bibfield  {journal}
  {\bibinfo  {journal} {arXiv e-prints}\ ,\ \bibinfo {eid} {arXiv:2006.06872}}
  (\bibinfo {year} {2020})},\ \Eprint {http://arxiv.org/abs/2006.06872}
  {arXiv:2006.06872 [hep-th]} \BibitemShut {NoStop}%
\bibitem [{\citenamefont {{Latora}}\ and\ \citenamefont
  {{Baranger}}(1999)}]{latora1999kolmogorov}%
  \BibitemOpen
  \bibfield  {author} {\bibinfo {author} {\bibfnamefont {Vito}\ \bibnamefont
  {{Latora}}}\ and\ \bibinfo {author} {\bibfnamefont {Michel}\ \bibnamefont
  {{Baranger}}},\ }\bibfield  {title} {\enquote {\bibinfo {title}
  {{Kolmogorov-Sinai Entropy Rate versus Physical Entropy}},}\ }\href {\doibase
  10.1103/PhysRevLett.82.520} {\bibfield  {journal} {\bibinfo  {journal}
  {\prl}\ }\textbf {\bibinfo {volume} {82}},\ \bibinfo {pages} {520--523}
  (\bibinfo {year} {1999})},\ \Eprint {http://arxiv.org/abs/chao-dyn/9806006}
  {arXiv:chao-dyn/9806006 [nlin.CD]} \BibitemShut {NoStop}%
\bibitem [{\citenamefont {{Nauenberg}}(2004)}]{nauenberg2004evolution}%
  \BibitemOpen
  \bibfield  {author} {\bibinfo {author} {\bibfnamefont {Michael}\ \bibnamefont
  {{Nauenberg}}},\ }\bibfield  {title} {\enquote {\bibinfo {title} {{The
  evolution of radiation toward thermal equilibrium: A soluble model that
  illustrates the foundations of statistical mechanics}},}\ }\href {\doibase
  10.1119/1.1632488} {\bibfield  {journal} {\bibinfo  {journal} {American
  Journal of Physics}\ }\textbf {\bibinfo {volume} {72}},\ \bibinfo {pages}
  {313--323} (\bibinfo {year} {2004})},\ \Eprint
  {http://arxiv.org/abs/cond-mat/0305219} {arXiv:cond-mat/0305219
  [cond-mat.stat-mech]} \BibitemShut {NoStop}%
\bibitem [{\citenamefont {{Kozlov}}\ and\ \citenamefont
  {{Treshchev}}(2007)}]{kozlov2007fine}%
  \BibitemOpen
  \bibfield  {author} {\bibinfo {author} {\bibfnamefont {V.~V.}\ \bibnamefont
  {{Kozlov}}}\ and\ \bibinfo {author} {\bibfnamefont {D.~V.}\ \bibnamefont
  {{Treshchev}}},\ }\bibfield  {title} {\enquote {\bibinfo {title}
  {{Fine-grained and coarse-grained entropy in problems of statistical
  mechanics}},}\ }\href {\doibase 10.1007/s11232-007-0040-1} {\bibfield
  {journal} {\bibinfo  {journal} {Theoretical and Mathematical Physics}\
  }\textbf {\bibinfo {volume} {151}},\ \bibinfo {pages} {539--555} (\bibinfo
  {year} {2007})}\BibitemShut {NoStop}%
\bibitem [{\citenamefont {Piftankin}\ and\ \citenamefont
  {Treschev}(2008)}]{piftankin2008gibbs}%
  \BibitemOpen
  \bibfield  {author} {\bibinfo {author} {\bibfnamefont {G.}~\bibnamefont
  {Piftankin}}\ and\ \bibinfo {author} {\bibfnamefont {D.}~\bibnamefont
  {Treschev}},\ }\bibfield  {title} {\enquote {\bibinfo {title} {Gibbs entropy
  and dynamics},}\ }\href {\doibase 10.1063/1.2907731} {\bibfield  {journal}
  {\bibinfo  {journal} {Chaos: An Interdisciplinary Journal of Nonlinear
  Science}\ }\textbf {\bibinfo {volume} {18}},\ \bibinfo {pages} {023116}
  (\bibinfo {year} {2008})},\ \Eprint
  {http://arxiv.org/abs/https://doi.org/10.1063/1.2907731}
  {https://doi.org/10.1063/1.2907731} \BibitemShut {NoStop}%
\bibitem [{\citenamefont {{{\v{Z}}upanovi{\'c}}}\ and\ \citenamefont
  {{Kui{\'c}}}(2018)}]{vzupanovic2018relation}%
  \BibitemOpen
  \bibfield  {author} {\bibinfo {author} {\bibfnamefont {Pa{\v{s}}ko}\
  \bibnamefont {{{\v{Z}}upanovi{\'c}}}}\ and\ \bibinfo {author} {\bibfnamefont
  {Domagoj}\ \bibnamefont {{Kui{\'c}}}},\ }\bibfield  {title} {\enquote
  {\bibinfo {title} {{Relation between Boltzmann and Gibbs entropy and example
  with multinomial distribution}},}\ }\href {\doibase 10.1088/2399-6528/aab7e1}
  {\bibfield  {journal} {\bibinfo  {journal} {Journal of Physics
  Communications}\ }\textbf {\bibinfo {volume} {2}},\ \bibinfo {eid} {045002}
  (\bibinfo {year} {2018})},\ \Eprint {http://arxiv.org/abs/1804.06818}
  {arXiv:1804.06818 [cond-mat.stat-mech]} \BibitemShut {NoStop}%
\bibitem [{\citenamefont {Gell-Mann}\ and\ \citenamefont
  {Hartle}(1993)}]{gellmann1993classical}%
  \BibitemOpen
  \bibfield  {author} {\bibinfo {author} {\bibfnamefont {Murray}\ \bibnamefont
  {Gell-Mann}}\ and\ \bibinfo {author} {\bibfnamefont {James~B.}\ \bibnamefont
  {Hartle}},\ }\bibfield  {title} {\enquote {\bibinfo {title} {Classical
  equations for quantum systems},}\ }\href {\doibase 10.1103/PhysRevD.47.3345}
  {\bibfield  {journal} {\bibinfo  {journal} {Phys. Rev. D}\ }\textbf {\bibinfo
  {volume} {47}},\ \bibinfo {pages} {3345--3382} (\bibinfo {year}
  {1993})}\BibitemShut {NoStop}%
\bibitem [{\citenamefont {{Dowker}}\ and\ \citenamefont
  {{Kent}}(1996)}]{dowker1996on}%
  \BibitemOpen
  \bibfield  {author} {\bibinfo {author} {\bibfnamefont {Fay}\ \bibnamefont
  {{Dowker}}}\ and\ \bibinfo {author} {\bibfnamefont {Adrian}\ \bibnamefont
  {{Kent}}},\ }\bibfield  {title} {\enquote {\bibinfo {title} {{On the
  consistent histories approach to quantum mechanics}},}\ }\href {\doibase
  10.1007/BF02183396} {\bibfield  {journal} {\bibinfo  {journal} {Journal of
  Statistical Physics}\ }\textbf {\bibinfo {volume} {82}},\ \bibinfo {pages}
  {1575--1646} (\bibinfo {year} {1996})},\ \Eprint
  {http://arxiv.org/abs/gr-qc/9412067} {arXiv:gr-qc/9412067 [gr-qc]}
  \BibitemShut {NoStop}%
\bibitem [{\citenamefont {Griffiths}(2019)}]{griffiths2019consistent}%
  \BibitemOpen
  \bibfield  {author} {\bibinfo {author} {\bibfnamefont {Robert~B.}\
  \bibnamefont {Griffiths}},\ }\bibfield  {title} {\enquote {\bibinfo {title}
  {The consistent histories approach to quantum mechanics},}\ }in\ \href@noop
  {} {\emph {\bibinfo {booktitle} {The Stanford Encyclopedia of Philosophy}}},\
  \bibinfo {editor} {edited by\ \bibinfo {editor} {\bibfnamefont {Edward~N.}\
  \bibnamefont {Zalta}}}\ (\bibinfo  {publisher} {Metaphysics Research Lab,
  Stanford University},\ \bibinfo {year} {2019})\ \bibinfo {edition} {summer
  2019}\ ed.\BibitemShut {Stop}%
\bibitem [{\citenamefont {Farmer}(1982)}]{farmer1982information}%
  \BibitemOpen
  \bibfield  {author} {\bibinfo {author} {\bibfnamefont {J.~Doyne}\
  \bibnamefont {Farmer}},\ }\bibfield  {title} {\enquote {\bibinfo {title}
  {Information dimension and the probabilistic structure of chaos},}\ }\href
  {\doibase https://doi.org/10.1515/zna-1982-1117} {\bibfield  {journal}
  {\bibinfo  {journal} {Zeitschrift für Naturforschung A}\ }\textbf {\bibinfo
  {volume} {37}},\ \bibinfo {pages} {1304 -- 1326} (\bibinfo {year}
  {1982})}\BibitemShut {NoStop}%
\bibitem [{\citenamefont {Frigg}(2004)}]{frigg2004sense}%
  \BibitemOpen
  \bibfield  {author} {\bibinfo {author} {\bibfnamefont {Roman}\ \bibnamefont
  {Frigg}},\ }\bibfield  {title} {\enquote {\bibinfo {title} {In what sense is
  the kolmogorov-sinai entropy a measure for chaotic behaviour?—bridging the
  gap between dynamical systems theory and communication theory},}\ }\href@noop
  {} {\bibfield  {journal} {\bibinfo  {journal} {The British journal for the
  philosophy of science}\ }\textbf {\bibinfo {volume} {55}},\ \bibinfo {pages}
  {411--434} (\bibinfo {year} {2004})}\BibitemShut {NoStop}%
\bibitem [{\citenamefont {Jost}(2006)}]{jost2006dynamical}%
  \BibitemOpen
  \bibfield  {author} {\bibinfo {author} {\bibfnamefont {J{\"u}rgen}\
  \bibnamefont {Jost}},\ }\href@noop {} {\emph {\bibinfo {title} {Dynamical
  systems: examples of complex behaviour}}}\ (\bibinfo  {publisher} {Springer
  Science \& Business Media},\ \bibinfo {year} {2006})\BibitemShut {NoStop}%
\bibitem [{\citenamefont {Daniel}(1984)}]{daniel1984entropy}%
  \BibitemOpen
  \bibfield  {author} {\bibinfo {author} {\bibfnamefont {W}~\bibnamefont
  {Daniel}},\ }\bibfield  {title} {\enquote {\bibinfo {title} {The entropy of
  observables on quantum logic},}\ }\href {\doibase
  10.1016/0034-4877(84)90004-1} {\bibfield  {journal} {\bibinfo  {journal}
  {Reports on mathematical physics}\ }\textbf {\bibinfo {volume} {19}},\
  \bibinfo {pages} {325--334} (\bibinfo {year} {1984})}\BibitemShut {NoStop}%
\bibitem [{\citenamefont {{Anz{\`a}}}\ and\ \citenamefont
  {{Vedral}}(2017)}]{anza2017information}%
  \BibitemOpen
  \bibfield  {author} {\bibinfo {author} {\bibfnamefont {F.}~\bibnamefont
  {{Anz{\`a}}}}\ and\ \bibinfo {author} {\bibfnamefont {V.}~\bibnamefont
  {{Vedral}}},\ }\bibfield  {title} {\enquote {\bibinfo {title}
  {{Information-theoretic equilibrium and observable thermalization}},}\ }\href
  {\doibase 10.1038/srep44066} {\bibfield  {journal} {\bibinfo  {journal}
  {Scientific Reports}\ }\textbf {\bibinfo {volume} {7}},\ \bibinfo {eid}
  {44066} (\bibinfo {year} {2017})}\BibitemShut {NoStop}%
\bibitem [{\citenamefont {{Lent}}(2019)}]{lent2019quantum}%
  \BibitemOpen
  \bibfield  {author} {\bibinfo {author} {\bibfnamefont {Craig~S.}\
  \bibnamefont {{Lent}}},\ }\bibfield  {title} {\enquote {\bibinfo {title}
  {{Quantum operator entropies under unitary evolution}},}\ }\href {\doibase
  10.1103/PhysRevE.100.012101} {\bibfield  {journal} {\bibinfo  {journal}
  {\pre}\ }\textbf {\bibinfo {volume} {100}},\ \bibinfo {eid} {012101}
  (\bibinfo {year} {2019})},\ \Eprint {http://arxiv.org/abs/1901.08956}
  {arXiv:1901.08956 [quant-ph]} \BibitemShut {NoStop}%
\bibitem [{\citenamefont {{Goldstein}}\ \emph {et~al.}(2019)\citenamefont
  {{Goldstein}}, \citenamefont {{Lebowitz}}, \citenamefont {{Tumulka}},\ and\
  \citenamefont {{Zanghi}}}]{goldstein2019gibbs}%
  \BibitemOpen
  \bibfield  {author} {\bibinfo {author} {\bibfnamefont {Sheldon}\ \bibnamefont
  {{Goldstein}}}, \bibinfo {author} {\bibfnamefont {Joel~L.}\ \bibnamefont
  {{Lebowitz}}}, \bibinfo {author} {\bibfnamefont {Roderich}\ \bibnamefont
  {{Tumulka}}}, \ and\ \bibinfo {author} {\bibfnamefont {Nino}\ \bibnamefont
  {{Zanghi}}},\ }\bibfield  {title} {\enquote {\bibinfo {title} {{Gibbs and
  Boltzmann Entropy in Classical and Quantum Mechanics}},}\ }\href@noop {}
  {\bibfield  {journal} {\bibinfo  {journal} {eprint}\ } (\bibinfo {year}
  {2019})},\ \Eprint {http://arxiv.org/abs/1903.11870} {arXiv:1903.11870
  [cond-mat.stat-mech]} \BibitemShut {NoStop}%
\bibitem [{\citenamefont {{Engelhardt}}\ and\ \citenamefont
  {{Wall}}(2019)}]{engelhardt2019coarse}%
  \BibitemOpen
  \bibfield  {author} {\bibinfo {author} {\bibfnamefont {Netta}\ \bibnamefont
  {{Engelhardt}}}\ and\ \bibinfo {author} {\bibfnamefont {Aron~C.}\
  \bibnamefont {{Wall}}},\ }\bibfield  {title} {\enquote {\bibinfo {title}
  {{Coarse graining holographic black holes}},}\ }\href {\doibase
  10.1007/JHEP05(2019)160} {\bibfield  {journal} {\bibinfo  {journal} {Journal
  of High Energy Physics}\ }\textbf {\bibinfo {volume} {2019}},\ \bibinfo {eid}
  {160} (\bibinfo {year} {2019})},\ \Eprint {http://arxiv.org/abs/1806.01281}
  {arXiv:1806.01281 [hep-th]} \BibitemShut {NoStop}%
\bibitem [{\citenamefont {Espa{\~n}ol}\ \emph {et~al.}(1997)\citenamefont
  {Espa{\~n}ol}, \citenamefont {Serrano},\ and\ \citenamefont
  {Zu{\~n}iga}}]{espanol1997coarse}%
  \BibitemOpen
  \bibfield  {author} {\bibinfo {author} {\bibfnamefont {Pep}\ \bibnamefont
  {Espa{\~n}ol}}, \bibinfo {author} {\bibfnamefont {Mar}\ \bibnamefont
  {Serrano}}, \ and\ \bibinfo {author} {\bibfnamefont {Ignacio}\ \bibnamefont
  {Zu{\~n}iga}},\ }\bibfield  {title} {\enquote {\bibinfo {title}
  {Coarse-graining of a fluid and its relation with dissipative particle
  dynamics and smoothed particle dynamic},}\ }\href {\doibase
  10.1142/S0129183197000771} {\bibfield  {journal} {\bibinfo  {journal} {Int.
  J. Mod. Phys. C}\ }\textbf {\bibinfo {volume} {08}},\ \bibinfo {pages}
  {899--908} (\bibinfo {year} {1997})},\ \Eprint
  {http://arxiv.org/abs/https://doi.org/10.1142/S0129183197000771}
  {https://doi.org/10.1142/S0129183197000771} \BibitemShut {NoStop}%
\bibitem [{\citenamefont {{Gao}}\ \emph {et~al.}(2017)\citenamefont {{Gao}},
  \citenamefont {{Betterton}}, \citenamefont {{Jhang}},\ and\ \citenamefont
  {{Shelley}}}]{gao2017analytical}%
  \BibitemOpen
  \bibfield  {author} {\bibinfo {author} {\bibfnamefont {Tong}\ \bibnamefont
  {{Gao}}}, \bibinfo {author} {\bibfnamefont {Meredith~D.}\ \bibnamefont
  {{Betterton}}}, \bibinfo {author} {\bibfnamefont {An-Sheng}\ \bibnamefont
  {{Jhang}}}, \ and\ \bibinfo {author} {\bibfnamefont {Michael~J.}\
  \bibnamefont {{Shelley}}},\ }\bibfield  {title} {\enquote {\bibinfo {title}
  {{Analytical structure, dynamics, and coarse graining of a kinetic model of
  an active fluid}},}\ }\href {\doibase 10.1103/PhysRevFluids.2.093302}
  {\bibfield  {journal} {\bibinfo  {journal} {Physical Review Fluids}\ }\textbf
  {\bibinfo {volume} {2}},\ \bibinfo {eid} {093302} (\bibinfo {year} {2017})},\
  \Eprint {http://arxiv.org/abs/1703.00969} {arXiv:1703.00969 [cond-mat.soft]}
  \BibitemShut {NoStop}%
\bibitem [{\citenamefont {Batchelor}\ and\ \citenamefont
  {Batchelor}(1967)}]{batchelor1967introduction}%
  \BibitemOpen
  \bibfield  {author} {\bibinfo {author} {\bibfnamefont {Cx~K}\ \bibnamefont
  {Batchelor}}\ and\ \bibinfo {author} {\bibfnamefont {GK}~\bibnamefont
  {Batchelor}},\ }\href@noop {} {\emph {\bibinfo {title} {An introduction to
  fluid dynamics}}}\ (\bibinfo  {publisher} {Cambridge university press},\
  \bibinfo {year} {1967})\BibitemShut {NoStop}%
\bibitem [{\citenamefont {Smith}(1950)}]{smith1950introduction}%
  \BibitemOpen
  \bibfield  {author} {\bibinfo {author} {\bibfnamefont {Joseph~Mauk}\
  \bibnamefont {Smith}},\ }\href@noop {} {\emph {\bibinfo {title} {Introduction
  to chemical engineering thermodynamics}}}\ (\bibinfo  {publisher} {ACS
  Publications},\ \bibinfo {year} {1950})\BibitemShut {NoStop}%
\bibitem [{\citenamefont {Guggenheim}(1956)}]{guggenheim1956statistical}%
  \BibitemOpen
  \bibfield  {author} {\bibinfo {author} {\bibfnamefont {E~A}\ \bibnamefont
  {Guggenheim}},\ }\href@noop {} {\emph {\bibinfo {title} {Statistical
  Thermodynamics; a Version of Statistical Mechanics for Students of Physics
  and Chemistry}}}\ (\bibinfo  {publisher} {The University press},\ \bibinfo
  {year} {1956})\BibitemShut {NoStop}%
\bibitem [{\citenamefont {Callen}(1998)}]{callen1998thermodynamics}%
  \BibitemOpen
  \bibfield  {author} {\bibinfo {author} {\bibfnamefont {Herbert~B}\
  \bibnamefont {Callen}},\ }\href@noop {} {\emph {\bibinfo {title}
  {Thermodynamics and an Introduction to Thermostatistics}}}\ (\bibinfo
  {publisher} {AAPT},\ \bibinfo {year} {1998})\BibitemShut {NoStop}%
\bibitem [{\citenamefont {Fisher}(1998)}]{fisher1998renormalization}%
  \BibitemOpen
  \bibfield  {author} {\bibinfo {author} {\bibfnamefont {Michael~E.}\
  \bibnamefont {Fisher}},\ }\bibfield  {title} {\enquote {\bibinfo {title}
  {Renormalization group theory: Its basis and formulation in statistical
  physics},}\ }\href {\doibase 10.1103/RevModPhys.70.653} {\bibfield  {journal}
  {\bibinfo  {journal} {Rev. Mod. Phys.}\ }\textbf {\bibinfo {volume} {70}},\
  \bibinfo {pages} {653--681} (\bibinfo {year} {1998})}\BibitemShut {NoStop}%
\bibitem [{\citenamefont
  {Kardar}(2007{\natexlab{a}})}]{kardar2007statisticalparticles}%
  \BibitemOpen
  \bibfield  {author} {\bibinfo {author} {\bibfnamefont {Mehran}\ \bibnamefont
  {Kardar}},\ }\href@noop {} {\emph {\bibinfo {title} {Statistical physics of
  particles}}}\ (\bibinfo  {publisher} {Cambridge University Press},\ \bibinfo
  {year} {2007})\BibitemShut {NoStop}%
\bibitem [{\citenamefont
  {Kardar}(2007{\natexlab{b}})}]{kardar2007statisticalfields}%
  \BibitemOpen
  \bibfield  {author} {\bibinfo {author} {\bibfnamefont {Mehran}\ \bibnamefont
  {Kardar}},\ }\href@noop {} {\emph {\bibinfo {title} {Statistical physics of
  fields}}}\ (\bibinfo  {publisher} {Cambridge University Press},\ \bibinfo
  {year} {2007})\BibitemShut {NoStop}%
\bibitem [{\citenamefont {Ma}(2018)}]{ma2018modern}%
  \BibitemOpen
  \bibfield  {author} {\bibinfo {author} {\bibfnamefont {Shang-Keng}\
  \bibnamefont {Ma}},\ }\href@noop {} {\emph {\bibinfo {title} {Modern theory
  of critical phenomena}}}\ (\bibinfo  {publisher} {Routledge},\ \bibinfo
  {year} {2018})\BibitemShut {NoStop}%
\bibitem [{\citenamefont {Wilson}(1971)}]{wilson1971renormalization}%
  \BibitemOpen
  \bibfield  {author} {\bibinfo {author} {\bibfnamefont {Kenneth~G.}\
  \bibnamefont {Wilson}},\ }\bibfield  {title} {\enquote {\bibinfo {title}
  {Renormalization group and critical phenomena. i. renormalization group and
  the kadanoff scaling picture},}\ }\href {\doibase 10.1103/PhysRevB.4.3174}
  {\bibfield  {journal} {\bibinfo  {journal} {\prb}\ }\textbf {\bibinfo
  {volume} {4}},\ \bibinfo {pages} {3174--3183} (\bibinfo {year}
  {1971})}\BibitemShut {NoStop}%
\bibitem [{\citenamefont
  {{\v{S}}afr{\'a}nek}(2020)}]{safranek2020observational}%
  \BibitemOpen
  \bibfield  {author} {\bibinfo {author} {\bibfnamefont {Dominik}\ \bibnamefont
  {{\v{S}}afr{\'a}nek}},\ }\bibfield  {title} {\enquote {\bibinfo {title}
  {Observational entropy with generalized measurements},}\ }\href@noop {}
  {\bibfield  {journal} {\bibinfo  {journal} {eprint}\ } (\bibinfo {year}
  {2020})},\ \Eprint {http://arxiv.org/abs/2007.07246} {arXiv:2007.07246
  [quant-ph]} \BibitemShut {NoStop}%
\bibitem [{\citenamefont {{Polkovnikov}}(2011)}]{polkovnikov2011microscopic}%
  \BibitemOpen
  \bibfield  {author} {\bibinfo {author} {\bibfnamefont {Anatoli}\ \bibnamefont
  {{Polkovnikov}}},\ }\bibfield  {title} {\enquote {\bibinfo {title}
  {{Microscopic diagonal entropy and its connection to basic thermodynamic
  relations}},}\ }\href {\doibase 10.1016/j.aop.2010.08.004} {\bibfield
  {journal} {\bibinfo  {journal} {Annals of Physics}\ }\textbf {\bibinfo
  {volume} {326}},\ \bibinfo {pages} {486--499} (\bibinfo {year} {2011})},\
  \Eprint {http://arxiv.org/abs/0806.2862} {arXiv:0806.2862
  [cond-mat.stat-mech]} \BibitemShut {NoStop}%
\bibitem [{\citenamefont {{von Neumann}}(2010)}]{vonNeumann1929}%
  \BibitemOpen
  \bibfield  {author} {\bibinfo {author} {\bibfnamefont {J.}~\bibnamefont {{von
  Neumann}}},\ }\bibfield  {title} {\enquote {\bibinfo {title} {{Proof of the
  ergodic theorem and the H-theorem in quantum mechanics. Translation of:
  Beweis des Ergodensatzes und des H-Theorems in der neuen Mechanik}},}\ }\href
  {\doibase 10.1140/epjh/e2010-00008-5} {\bibfield  {journal} {\bibinfo
  {journal} {Eur. Phys. J. H}\ }\textbf {\bibinfo {volume} {35}},\ \bibinfo
  {pages} {201--237} (\bibinfo {year} {2010})},\ \Eprint
  {http://arxiv.org/abs/1003.2133} {arXiv:1003.2133 [physics.hist-ph]}
  \BibitemShut {NoStop}%
\end{thebibliography}%

\appendix

%\section{Table of comparisons}

\section{Physically relevant coarse-grainings}

In this appendix we collect and/or introduce, and discuss, a number of physically relevant coarse-grainings, generally with some relevance to thermodynamics, some of which have been studied in detail before~\cite{vonNeumann1929,safranek2019short,safranek2019long,safranek2019classical,strasberg2019entropy,strasberg2020heat}.

In terms of generic properties, any observational entropy generated by an observable (which is typically the case of {\bf (1)} below) that is conserved in the system (i.e., commutes with the Hamiltonian) will be constant during the time evolution~\cite{safranek2019long}. Moreover, any entropy that consists solely of local coarse-grainings (which is the case of {\bf (2)} below) will be additive on independent systems as per Eq.~\eqref{eq:additive}, and bounded as per~\eqref{eq:local_bound}. %physical interpretation of local cg entropy
All convergence in the long-time limit discussed below in points {\bf (2)} and {\bf (3)} assumes particle conserving non-integrable\footnote{Although simulations~\cite{safranek2019long} show that integrable case also converges, just not that well.} Hamiltonian with short range interactions, so that particles tend to thermalize well and the interaction energy between the subsystems is negligible. The convergence holds up to some corrections due to finite particle number and finite size-effect effects. These limits are approximate upper bounds for the non-equilibrium entropies, which is why we say ``grows to.'' The exact upper bounds follow from Eq.~\eqref{eq:nonincreasing}. For simplicity we also consider non-degenerate Hamiltonian (both globally and locally), which means that each energy has a unique associated particle number, so instead of common eigenstate $\ket{n,E}$ of the particle operator and the Hamiltonian we can write simply $\ket{E}$. The two observational entropies discussed in the main body of this paper are {\bf (1c)} and {\bf (2c)}.

\emph{\bf (1a) Global particle number coarse-graining}\\
Defining $\C_{\hat{N}}=\{\P_n\}$, where $\P_n$ is a projector onto subspace of $n$ particles,
\[
S_{\C_{\hat{N}}}=-\sum_np_n\ln \frac{p_{n}}{V_{n}}
\]
measures the uncertainty about the particle number in the system.\\

\emph{\bf (1b) Global energy coarse-graining}\footnote{This has been originally defined by von Neumann in~\cite{vonNeumann1929}, where he attributed this definition to Eugene Wigner. It was used extensively in both classical~\cite{safranek2019classical} and quantum case~\cite{safranek2019short,safranek2019long}, usually as the value to which other entropies $S_{xE}$ and $S_F$ (in case of fixed total number of particles) converge.}\\
Defining $\C_E\equiv C_{\hat{H}^{(\DE)}}=\{\P_E\}$, where $\P_E=\sum_{E\leq \tilde E <E+\DE}\pro{\tilde E}{\tilde E}$ is a projector onto subspace of wave functions within an energy shell $[E,E+\DE)$ (and $\hat{H}^{(\DE)}=\sum_E E \P_E$ is the coarse-grained Hamiltonian),
\[
S_{\C_E}=-\sum_E p_E \ln \frac{p_E}{V_E}
\]
measures the equilibrium thermodynamic entropy of a system with a fixed number number of particles.

\emph{Details:}
$\DE$ is the resolution in energy of the measuring apparatus. If restricted to a Hilbert space with a fixed number of particles, for small but non-zero $\DE$ this entropy gives microcanonical entropy for both energy eigenstates and a microcanonical state, and it (approximately) gives Gibbs entropy $\ln Z-\beta\mean{E}$ for the Gibbs state $\frac{1}{Z}e^{-\beta \hat{H}}$. The case of $\DE>0$ cannot be applied to Hilbert space which includes variable number of particles, because the energy subspace would include wave-functions with any particle numbers, and would be typically infinite in size. For  $\DE=0$ it does not have this problem (since energy eigenstate uniquely determines the particle number in common particle-conserving Hamiltonians), but it gives zero for energy eigenstates.

\emph{{\bf (1c)} \bf Global particle number with global energy coarse-graining}\\
\[
S_{\th}\equiv S_{\C_{\hat{N}},\C_E}=-\sum_{n,E} p_{nE} \ln \frac{p_{nE}}{V_{nE}}
\]
measures the equilibrium thermodynamic entropy.

\emph{Details:} Macrostates now distinguish both energy and the number of particles, which means it can be used for systems with a variable number of particles. For common particle-conserving Hamiltonians, the case of $\DE=0$ reduces to $S_{\C_E}$. It gives $S_{\C_{\hat{N}},\C_E}(\ket{E})=\ln V_{nE}=S_{\mathrm{micro}}(E,\mathcal{V},n)$ microcanonical entropy, for a global energy eigenstate. $\mathcal{V}$ is the spatial volume of the system.

\emph{\bf (2a) Local particle number coarse-graining}\footnote{Denoted $S_x$ in~\cite{safranek2019long} where its time evolution is illustrated in Fig. 2.}\\
\[
S_{\C_{\hat{N}_1}\otimes\cdots\otimes \C_{\hat{N}_m}}=-\sum_\bn p_\bn\ln \frac{p_{\bn}}{V_{\bn}}
\]
where $\bn=(n_1,\dots,n_m)$ are energies of the subsystems, and measures how uniformly are particles distributed over the subsystems.

\emph{Details:} It grows to (1a) in the long-time limit: when particles spread uniformly throughout the system, they fill uniformly every particle shell.

\emph{\bf (2b) Local energy coarse-graining}\footnote{The case of $\DE=0$ has been studied in detail in~\cite{safranek2019long} under the name of ``Factorized Observational entropy'' or FOE for short, and denoted $S_F$.}\\
\[
S_{\C_{E_1}\otimes\cdots\otimes \C_{E_m}}=-\sum_\bE p_\bE\ln \frac{p_{\bE}}{V_{\bE}}
\]
where $\bE=(E_1,\dots,E_m)$ are energies in the subsystems, how uniformly is energy distributed over the subsystems.

\emph{Details:} For local energy eigenstates $S_{\C_{E_1}\otimes\cdots\otimes \C_{E_m}}(\ket{\tilde E_1}\cdots\ket{\tilde E_m})=\sum_{i=1}^m\ln V_{\tilde E_i}$. Despite from what it may seem from {(1b)}, for $\DE>0$, $\ln V_{\tilde E_i}$ does not describe thermodynamic entropy in each subsystem, because in a non-equilibrium system, number of particles in each subsystem typically varies, even though the total number of particles may be conserved. Macrostate $\HS_{E_i}$ contains all states with energy $E_i$, even though these states might have different particle numbers. In case of Hamiltonians which conserve particles locally, each local eigenstate uniquely determines its particle number, which implies that the case of $\DE=0$ is identical to {(2c)}, having all of its dynamical properties. However, for $\DE=0$, $S_{\C_{E_1}\otimes\cdots\otimes \C_{E_m}}(\ket{\tilde E_1}\cdots\ket{\tilde E_m})=0$, which is undesirable for a physically meaningful thermodynamic entropy. It grows to (1b) in the long-time limit.

\emph{\bf (2c) Local particle number with local energy coarse-graining}\footnote{This entropy has been studied closely in the classical case~\cite{safranek2019classical} where it has been denoted $S_F$, and where also the quantum equivalent is mentioned for the first time. Since in quantum case, this definition behaves the same (in its time evolution in particular) as (2b) for $\DE=0$, apart from giving non-zero value for local energy eigenstates, we refer reader to~\cite{safranek2019long} for its detailed properties.}\\
\[
S_{\th}^{\mathrm{non-eq.}}\equiv S_{\C_{\hat{N}_1}\otimes\cdots\otimes \C_{\hat{N}_m},\C_{E_1}\otimes\cdots\otimes \C_{E_m}}=-\sum_{\bn,\bE} p_{\bn\bE}\ln \frac{p_{\bn\bE}}{V_{\bn\bE}}
\]
measures non-equilibrium thermodynamic entropy of the system.

\emph{Details:} At some intermediate time~$t$ (when the system has only partially equilibrated)  
its value can be interpreted as the equilibrium thermodynamic entropy the system would attain in the long-time limit if (hypothetically) starting from time $t$ the subsystems were not allowed to exchange either energy or particles~\cite{safranek2019classical}. For $\DE>0$, it gives $S_{\C_{\hat{N}_1}\otimes\cdots\otimes \C_{\hat{N}_m},\C_{E_1}\otimes\cdots\otimes \C_{E_m}}(\ket{E_1}\cdots\ket{E_m})=\sum_{i=1}^m\ln V_{nE}=\sum_{i=1}^mS_{\mathrm{micro}}(E_i,\mathcal{V}_i,n_i)$, the sum of local microcanonical entropies, for a local energy eigenstate. $\mathcal{V}_i$ denote the local spatial volumes. It grows to (1c) in the long-time limit.

\emph{\bf (3a) Local particle number then global energy coarse-graining}\footnote{This entropy has been studied in detail both in classical~\cite{safranek2019classical} and quantum case~\cite{safranek2019long} where it was denoted $S_{xE}$.}\\
\[
S_{\C_{\hat{N}_1}\otimes\cdots\otimes \C_{\hat{N}_m},\C_{E}}=-\sum_{\bn,E} p_{\bn E}\ln \frac{p_{\bn E}}{V_{\bn E}}
\]
is a different type of non-equilibrium thermodynamic entropy of the system.

\emph{Details:} Is not additive. At some intermediate time~$t$, its value can be interpreted as the equilibrium thermodynamic entropy the system would attain in the long-time limit if (hypothetically) starting from time $t$ the subsystems were allowed to exchange energy but not particles~\cite{safranek2019classical}. It is upper bounded by (2a), and it grows to (1c).

\emph{\bf (3b) Global energy then local particle number coarse-graining}\footnote{Time evolution of this entropy has been studied in the quantum case in Appendix~H of~\cite{safranek2019long}, where it was denoted $S_{Ex}$, but it was realized only later in~\cite{safranek2019classical} that it has a meaningful interpretation. Classically, $S_{xE}$ and $S_{Ex}$ are identical.}\\
\[
S_{\C_{E},\C_{\hat{N}_1}\otimes\cdots\otimes \C_{\hat{N}_m}}=-\sum_{E,\bn} p_{E\bn}\ln \frac{p_{ E\bn}}{V_{E\bn}}
\]
is similar in behavior to (3a), but differs when quantum effects become significant, such as at low energies and when subsystems are small so that effects of non-commutation between $\hat{N_i}$ and $\hat{H}$ intensify.

\emph{Details:} It is upper bounded by (1b), and it grows to (1c). For $\DE=0$, it is identical to (1b) and (1c).

\emph{\bf (4) Combination of arbitrary local and local energy coarse-graining.}\footnote{This entropy has been studied in detail by Strasberg and Winter in~\cite{strasberg2019entropy,strasberg2020heat}.}\\
\[
S_{\C\otimes\C_{\hat{E}_1}\otimes\cdots\otimes \C_{E_m}}=-\sum_{i, \bE} p_{i \bE}\ln \frac{p_{i \bE}}{V_{i \bE}}
\]
is the total entropy of a small well-controlled subsystem plus large bath(s), with applications in open system non-equilibrium thermodynamics.

\emph{Details:} Change in this entropy defines entropy production which, unlike the formulation based on von Neumann entropy, does not depend explicitly on the temperature(s) of the bath(s). Moreover, with this it is possible to define work as the part of the ``useful'' internal energy that can be recovered from the system, while heat is the part of internal energy that is irreversibly lost.
%The treatment thus far has pertained to any possible coarse-graining. However, even if entropy increase is generic, there is no reason to expect that an arbitrary coarse-graining will be closely connected with {\em thermodynamics}, which in particular relates temperature, energy and entropy.\\

\end{document}